\documentclass[manuscript,screen]{acmart}
\usepackage{hyperref}  % Alternative that includes URL support
\usepackage{listings}  % For code listings
\usepackage{color}
\definecolor{lightgray}{rgb}{.9,.9,.9}
\definecolor{darkgray}{rgb}{.4,.4,.4}
\definecolor{purple}{rgb}{0.65, 0.12, 0.82}

\lstdefinelanguage{JavaScript}{
  keywords={typeof, new, true, false, catch, function, return, null, catch, switch, var, if, in, while, do, else, case, break},
  keywordstyle=\color{blue}\bfseries,
  ndkeywords={class, export, boolean, throw, implements, import, this},
  ndkeywordstyle=\color{darkgray}\bfseries,
  identifierstyle=\color{black},
  sensitive=false,
  comment=[l]{//},
  morecomment=[s]{/*}{*/},
  commentstyle=\color{purple}\ttfamily,
  stringstyle=\color{red}\ttfamily,
  morestring=[b]',
  morestring=[b]"
}

\AtBeginDocument{%
  }

\setcopyright{acmlicensed}
\copyrightyear{2025}
\acmYear{2025}
\acmDOI{XXXXXXX.XXXXXXX}
\acmJournal{TOPS}
\acmVolume{0}
\acmNumber{0}
\acmArticle{0}

\begin{document}

\title{interID - An Ecosystem-agnostic Verifier-as-a-Service with OpenID Connect Bridge}


\author{Hakan Yildiz}
\affiliation{%
  \institution{Service-centric Networking, Technische Universität Berlin}
  \city{Berlin}
  \country{Germany}}
\email{hakan@interid.io}
\orcid{0000-0002-2044-0826}

\author{Axel Küpper}
\affiliation{%
  \institution{Service-centric Networking, Technische Universität Berlin}
  \city{Berlin}
  \country{Germany}}
\email{axel.kuepper@tu-berlin.de}
\orcid{00000-0002-4356-5613}

\renewcommand{\shortauthors}{Yildiz and Küpper}

\begin{abstract}
Self-Sovereign Identity (SSI) addresses fundamental limitations of traditional
identity systems through user-controlled, cryptographically verifiable credentials stored in wallet applications.
As European Union regulations mandate EUDI Wallet acceptance by public and private
sector organizations for specific use cases by 2027, SSI adoption transitions from optional to required.
However, integration complexity presents a critical barrier: each SSI Verifier application
exposes different APIs with distinct request parameters, response formats, and claim structures,
requiring custom integration wrappers and dedicated infrastructure deployment. This
contrasts with federated identity systems where OpenID Connect (OIDC) combined with
cloud-based providers enables seamless integration without infrastructure management.

interID is an ecosystem-agnostic verification platform that unifies credential
verification across Hyperledger Aries/Indy, EBSI, and EUDI ecosystems through a
single orchestration layer. We extend interID with an OIDC bridge that provides a
Verifier-as-a-Service offering, enabling SSI verification through standardized
authentication protocols. Organizations integrate SSI-based authentication using
standard OIDC flows without deploying verification infrastructure, receiving ID Tokens
with cryptographically verified credential attributes without implementing
Verifier-specific or credential-format-specific logic. The multi-tenant architecture
leverages Keycloak for Identity and Access Management (IAM) with strict tenant isolation,
enabling shared infrastructure while maintaining independent organizational
configurations. Key innovations include explicit PKCE support, flexible
scope-to-proof-template mappings that translate standard OIDC scopes into
ecosystem-specific verification requests, and a security analysis identifying
novel attack surfaces that emerge at the intersection of OIDC, SSI, and
multi-tenant architectures---threats addressed by neither RFC 6819 (OAuth 2.0
threat model) nor existing SSI security analyses in isolation.

Our evaluation demonstrates security equivalence to production identity providers
through threat modeling identifying 11 attack vectors with defense-in-depth
countermeasures, including seven threats beyond RFC 6819's scope spanning
SSI-specific vulnerabilities and multi-tenant isolation violations. Integration complexity analysis using
representative use cases shows that organizations can adopt SSI-based authentication
with comparable effort to adding traditional federated identity providers, leveraging
existing OIDC infrastructure without deploying or managing verification services. By combining
familiar OIDC integration patterns with SaaS deployment, our work lowers both
integration and operational barriers, enabling organizations to meet regulatory
requirements through configuration rather than custom development and infrastructure
deployment.
\end{abstract}




%% Received dates will be set by ACM upon submission
%%\received{Date TBD}
%%\received[revised]{Date TBD}
%%\received[accepted]{Date TBD}

\maketitle

\section{Introduction}

Self-Sovereign Identity (SSI) represents a fundamental shift in digital identity management, empowering individuals with direct control over their credentials without reliance on continuous access to Identity Providers (IdPs). By enabling users to receive cryptographically verifiable credentials from trusted issuers, store them in digital wallets, and present them selectively to verifying organizations, SSI addresses fundamental limitations of traditional identity systems, including restricted credential portability, privacy concerns, limited credential reusability, and constrained user autonomy~\cite{yildiz2023interoperable}. The promise of SSI has catalyzed the development of diverse technical frameworks including Hyperledger Indy\footnote{Hyperledger Indy: \url{https://lf-hyperledger.atlassian.net/wiki/spaces/indy/overview}}/Aries\footnote{Hyperledger Aries: \url{https://lf-hyperledger.atlassian.net/wiki/spaces/ARIES/overview}}, the European Blockchain Services Infrastructure (EBSI)\footnote{European Blockchain Services Infrastructure: \url{https://ec.europa.eu/digital-building-blocks/sites/display/EBSI/Home}}, and the EU Digital Identity (EUDI)\footnote{European Digital Identity: \url{https://commission.europa.eu/strategy-and-policy/priorities-2019-2024/europe-fit-digital-age/european-digital-identity_en}}, each offering distinct approaches to credential formats, exchange protocols, and trust models~\cite{yildiz_interop_2022}.

Despite this progress, SSI adoption faces a critical dual challenge for organizations seeking to integrate credential verification capabilities. First, while our previous work on interID successfully addressed cross-ecosystem credential verification through a unified orchestration layer~\cite{yildiz2025}, each Verifier application, including interID, exposes different APIs to integrating organizations, each with distinct request parameters, response formats, and claim structures. For example, initiating credential verification requires \texttt{POST /ui/presentations} in the EUDI Reference Implementation but \texttt{POST /api/proof-request/v1/:id/share} in Procivis One Core---different endpoints, parameters, and response schemas despite both supporting the same EUDI ecosystem. This forces organizations to implement custom integration logic despite standardization efforts at the wallet-to-Verifier interface. Second, traditional Verifier deployments require organizations to host and maintain ecosystem-specific verification infrastructure, imposing substantial operational overhead. Together, these barriers compel organizations to develop custom backend implementations for each Verifier integration, requiring specialized SSI knowledge and ongoing maintenance of both integration code and infrastructure.

These integration challenges parallel those that existed in traditional federated identity systems before the emergence of standardized protocols. OAuth 2.0 and OpenID Connect (OIDC) solved precisely these problems, with OIDC becoming the de facto standard for authentication and authorization of natural persons~\cite{curity2023oauth}. OIDC's success stems from providing a uniform northbound interface and standardized JSON Web Token (JWT)-based ID Tokens containing identity information. Consequently, integrating a new IdP requires minimal effort. Organizations simply configure standardized endpoints and consume ID Tokens, regardless of the underlying IdP's internal mechanisms. This standardization enabled rapid adoption of diverse IdPs including Microsoft Entra ID, Google Identity~\cite{fett2017web}, Okta, Auth0, and social login platforms, leveraging familiar integration patterns that developers have implemented repeatedly.

Current SSI authentication approaches, however, fail to provide equivalent standardization. Beyond the lack of uniform northbound interfaces, SSI faces an additional complexity: even when using protocols like OpenID for Verifiable Presentations (OID4VP)~\cite{oid4vp} and Self-Issued OpenID Provider v2 (SIOPv2)~\cite{siopv2} that bring OAuth-based flows to SSI, verification results remain fundamentally heterogeneous across credential formats. AnonCreds zero-knowledge proofs, JSON-LD linked data structures, and ISO mDoc CBOR encodings each require format-specific parsing, cryptographic verification, and claim extraction mechanisms~\cite{yildiz_interop_2022}. This dual challenge forces organizations to implement both Verifier-application-specific and credential-format-specific integration logic, creating complexity that far exceeds traditional federated identity integration despite progress in protocol standardization. While several commercial SSI platforms offer Software-as-a-Service (SaaS) deployment models, these platforms similarly expose vendor-specific APIs and response formats, requiring custom integration development for each provider rather than leveraging standardized authentication patterns familiar to developers.

The urgency of addressing these challenges has increased substantially with the European Union's revised eIDAS regulation, which mandates that member states issue EUDI Wallets to citizens, with organizations required to accept EUDI Wallet credentials for specific use cases by November 2027~\cite{eidas2}.\footnote{The supporting implementing acts for eIDAS 2.0 entered into force in November 2024, establishing November 2027 as the compliance deadline.} This regulatory requirement transforms SSI adoption from voluntary exploration into a compliance necessity, affecting use cases ranging from Know Your Customer (KYC) identity proofing in financial services to single sign-on across enterprise application portfolios. For organizations requiring support for multiple SSI ecosystems---EUDI for European operations, Aries/Indy for North American or Asian partnerships,\footnote{Hyperledger Indy/Aries is predominantly deployed in Canada and Bhutan, though Bhutan is currently migrating to PolygonID.} and EBSI for cross-border services---integration costs multiply substantially as each ecosystem requires independent implementation with limited code reuse, contrasting sharply with federated identity adoption where standardized OIDC protocols combined with SaaS delivery models enabled rapid, low-effort integration.

To address this dual challenge while enabling organizations to meet regulatory deadlines and control costs, we propose an OIDC bridge that provides the missing standardized northbound interface and SaaS deployment model for SSI authentication. Our approach extends interID's ecosystem-agnostic verification platform with an OIDC-compliant authentication server delivered as a multi-tenant SaaS offering, abstracting both Verifier API diversity and credential format heterogeneity while eliminating infrastructure deployment requirements. Organizations integrate SSI-based authentication using standard OIDC authorization requests with existing client libraries, receiving ID Tokens containing cryptographically verified credential attributes without implementing SSI-specific logic, deploying verification services, or managing infrastructure. This approach mirrors the successful integration pattern established by federated identity providers, where organizations add new authentication mechanisms through configuration rather than custom development. Just as developers routinely add social login options by configuring standard OIDC endpoints, they can now integrate SSI-based authentication using the same familiar patterns, reducing both implementation effort and total cost of ownership.

The primary contributions of this work include:

\begin{itemize}
    \item An OIDC-compliant authentication server that bridges SSI credential verification with standardized authentication protocols, enabling organizations to integrate SSI through standard OIDC flows without implementing Verifier-specific wrappers, credential-format-specific parsing logic, or cryptographic verification mechanisms, and without deploying or managing verification infrastructure.
    
    \item A multi-tenant SaaS architecture leveraging Keycloak for client management and supporting cloud deployment with proper tenant isolation. This architectural approach enables multiple organizations to share infrastructure while maintaining independent configurations, reducing operational overhead and deployment costs compared to traditional on-premises Verifier deployments.
    
    \item A scope-to-proof-template mapping mechanism that translates standard OIDC scopes into ecosystem-specific verification requests, enabling organizations to define verification requirements through OIDC configuration rather than SSI-specific logic. The complete implementation supports Hyperledger Aries/Indy, EBSI, and EUDI ecosystems with Authorization Code Flow with Proof Key for Code Exchange (PKCE), session management across OIDC and SSI verification flows, and secure ID Token-based result transmission.

    \item A security analysis identifying seven novel attack threats beyond RFC 6819's OAuth 2.0 threat model, emerging at the intersection of SSI credential verification and multi-tenant OIDC bridge architectures. These include SSI-specific threats (proof request manipulation, verification result spoofing, credential presentation replay) and SaaS-specific multi-tenant isolation threats not present in single-deployment models, addressed through nine defense-in-depth security controls.
    
    \item An evaluation demonstrating integration convenience through qualitative comparison with direct SSI integration using representative use cases, validated by a reference Relying Party implementation requiring approximately 180 lines of standard OIDC client code with no SSI-specific dependencies, and validation of OIDC protocol compliance enabling integration through standard authorization flows and token exchange mechanisms.
\end{itemize}

By eliminating Verifier-specific API integration, credential format parsing, and operational infrastructure requirements, our work enables organizations to adopt SSI-based authentication through standard OIDC flows. Organizations leverage the same OIDC authorization flows, token endpoints, and ID Token parsing they use for existing federated IdPs, maintaining standard OIDC protocol flows, token formats, and security properties. This represents a crucial step toward mainstream SSI adoption, enabling organizations to meet regulatory compliance deadlines while substantially reducing implementation costs and maintaining the flexibility to evolve their SSI strategies as the ecosystem matures.

The remainder of this paper is organized as follows. Section~\ref{background} provides essential background on traditional identity paradigms, the OIDC protocol, SSI authentication challenges, and our previous work, culminating in a precise problem statement. Section~\ref{related} surveys related work and identifies our research gap. Section~\ref{design} presents our architectural design. Section~\ref{implementation} details the implementation. Section~\ref{evaluation} evaluates our solution. Section~\ref{future} discusses future work, and Section~\ref{conclusion} concludes this work.

\section{Background and Problem Statement} \label{background}

Understanding why SSI authentication faces integration challenges requires examining how traditional identity systems solved similar problems through protocol standardization. The evolution from centralized to federated identity systems encountered the same fundamental challenge we observe in SSI today: diverse implementations with different northbound interfaces created integration barriers that hindered adoption. OAuth 2.0 and OpenID Connect emerged as solutions to this problem, establishing standardized protocols that enabled seamless integration regardless of provider choice. By analyzing the history of digital identity paradigms and OIDC's standardization approach, particularly its uniform northbound interface, we can identify precisely where SSI authentication deviates from this proven pattern. 

\subsection{Architectural Interface Terminology}

We adopt terminology from Software-Defined Networking to describe system layer communication. A \textit{northbound interface} enables components to communicate with higher-level applications, while a \textit{southbound interface} establishes communication with lower-level services.

In identity verification systems: the \textbf{southbound interface} handles credential presentation between wallets and Verifier applications (using DIDComm for Aries, OID4VP for EBSI/EUDI). The \textbf{northbound interface} delivers verification results from Verifier applications to organizations' backend systems. While individual SSI ecosystems have standardized southbound interfaces for wallet-to-Verifier communication (DIDComm for Aries, OID4VP for EBSI/EUDI), northbound interfaces remain fragmented as each Verifier application exposes different APIs with distinct formats, forcing organizations to implement custom integration logic.

\subsection{Traditional Identity Paradigms}

Digital identity systems have evolved through three distinct paradigms, each with different trust models and control mechanisms. In \textit{centralized identity} systems, a single authority stores and manages all user credentials. Users authenticate by providing a username and password directly to the organization providing services, which validates these credentials against its own identity database. This model offers simplicity but suffers from scalability limitations, security vulnerabilities associated with centralized credential storage, and poor credential reusability across different organizations.

\textit{Federated identity} systems address these limitations by separating authentication from service provision through trusted third parties. In this model, IdPs authenticate users and issue tokens or assertions to organizations that rely on this identity information. Users authenticate once with their IdP and gain access to multiple organizations and their constituent services through trust relationships. This paradigm reduces password proliferation and enables single sign-on (SSO) experiences familiar to enterprise users through platforms like Microsoft Entra ID, Google Identity, and Okta. 

These \textit{Identity-as-a-Service} offerings have become ubiquitous in modern web applications, with OIDC emerging as the dominant federation protocol for authentication of natural persons. 

\subsection{OpenID Connect Protocol}

OIDC is an identity layer built on OAuth 2.0 that standardizes authentication
through a consistent set of endpoints and token formats~\cite{siriwardena2019openid,
oidccore}. The protocol defines OpenID Providers (OPs) that issue identity tokens
and Relying Parties (RPs) that consume them. OIDC supports multiple authentication flows---including the implicit flow, the authorization code flow, and hybrid combinations---each suited to different client types and security requirements. OIDC's success stems from providing
uniform interfaces: regardless of whether an organization integrates
Microsoft Entra ID, Google Identity, or Okta, they use identical discovery endpoints
(\texttt{.well-known/openid-configuration}), authorization endpoints, token endpoints,
and JWKS endpoints for signature validation. This standardization enables RPs to
integrate new IdPs through configuration rather than custom development.

\subsubsection{Authorization Code Flow with PKCE}\label{sec:pkce-background}

\begin{figure}[!htbp]
  \centering
  \includegraphics[width=\columnwidth]{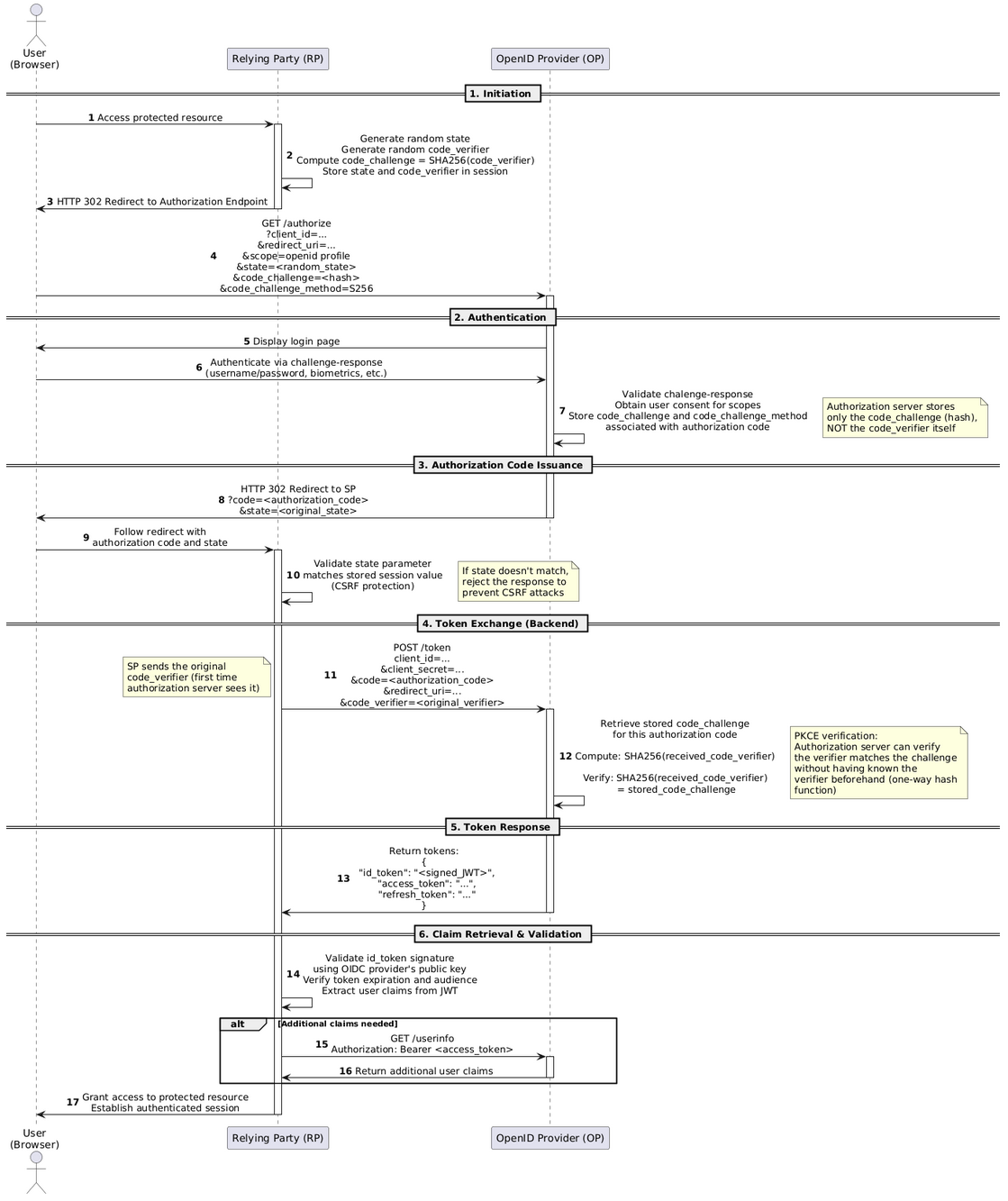}
  \caption{Visualization of Authorization Code Flow with Proof Key for Code Exchange}
  \Description{Sequence diagram showing the complete OIDC Authorization Code Flow with PKCE, illustrating the six phases of authentication including initiation, user authentication, authorization code issuance, backend token exchange with PKCE verification, token response, and claim retrieval}
  \label{fig:oidc-pkce}
\end{figure}

Among these flows, the authorization code flow with Proof Key for Code Exchange (PKCE) provides the
strongest security guarantees for OIDC authentication~\cite{rfc6749} and forms the basis for our bridge implementation. As illustrated 
in Figure~\ref{fig:oidc-pkce}, the flow cryptographically binds authorization 
requests to token exchanges through challenge-response mechanisms: the RP generates 
a \texttt{code\_verifier}, derives a \texttt{code\_challenge}, and includes it in 
the authorization request. The OP stores the challenge and later validates that the 
\texttt{code\_verifier} presented during token exchange matches, preventing 
authorization code interception attacks. This protection is crucial for public 
clients (single-page applications, mobile apps) that cannot securely store client 
secrets, but also provides defense-in-depth for confidential clients.

\subsubsection{Why OIDC Succeeds: Integration Advantages}

For organizations implementing authentication, OIDC integration offers compelling advantages. The protocol is standardized by the OpenID Foundation with extensive specifications and interoperability testing. Well-maintained libraries exist for all major programming languages and frameworks, significantly reducing implementation complexity. Developers familiar with OAuth 2.0 can leverage existing knowledge, as OIDC extends rather than replaces OAuth. The redirect-based flow integrates seamlessly with web application architectures, requiring minimal changes to existing authentication systems. Importantly, the RP only implements a standard OIDC client: it requires no understanding of the IdP's internal authentication mechanisms, credential formats, or storage systems.

\subsection{Self-Sovereign Identity}
While OIDC and federated identity models solve substantial integration problems securely, federated identity systems (along with earlier centralized models) share fundamental limitations: users lack direct control over their identity data, credentials are not portable across federation boundaries, and privacy concerns arise from centralized tracking of user authentication activities. The promises of SSI aim to address these challenges.

In the SSI model, IdPs evolve into Issuers who create and cryptographically sign Verifiable Credentials (VCs). These VCs are issued to Identity Holders (typically end users). Identity Holders store VCs in a wallet application on their personal devices, e.g., on their smartphones. Identity Holders present them to Verifiers (RP equivalent in SSI terminology) as Verifiable Presentations (VPs). VPs contain critical assurances, including proof of credential ownership and non-revocation, and incorporate features like audience restriction and nonce mechanisms to mitigate replay attacks. Furthermore, verification relies on trust registries, also known as Verifiable Data Registries, that contain the relevant data needed to validate the integrity and authenticity of credentials, without requiring direct communication between the Issuers and the Verifiers~\cite{yildiz2025}.
%self cite interID paper

\begin{figure}[!htbp]
  \centering
  \includegraphics[width=\columnwidth]{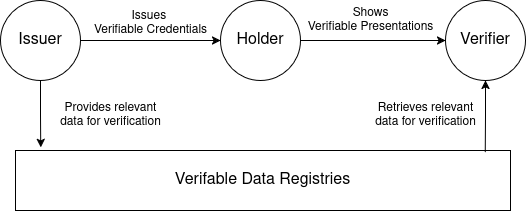}
  \caption{High-level SSI model}
  \Description{Diagram showing the three main roles and their interactions in SSI: Issuer, Holder, and Verifier while depicting the role of a Verifiable Data Registry}
  \label{fig:ssi_roles}
\end{figure}

Figure~\ref{fig:ssi_roles} depicts the interactions between Issuers, Identity Holders, and Verifiers as well as the role of Verifiable Data Registries.

\subsubsection{SSI Authentication Approaches: The EUDI Example}

Several approaches to SSI-based authentication have emerged, each with different integration characteristics. As a representative example, the EUDI Architecture and Reference Framework specifies user identification and authentication using Person Identification Data (PID) exchanged through OID4VP for remote credential presentation. After initial identification with a PID, services can issue authentication attestations to the wallet, which users can subsequently present for accessing services~\cite{eudi_arf_1.6}.

However, these authentication mechanisms remain ecosystem-specific. The EUDI approach works well within the European Digital Identity ecosystem but requires Verifiers to implement EUDI-specific protocols and profile configurations. Similarly, Hyperledger Aries-based authentication requires understanding of DIDComm protocols and different credential formats. Each ecosystem provides its own authentication flow with specific credential formats and verification mechanisms.

\subsubsection{The Integration Challenge}

While protocols like OID4VP and SIOPv2 bring OAuth-based flows to SSI for wallet-to-Verifier communication, the northbound integration challenge persists. As illustrated by the API endpoint differences noted in Section~1 (e.g., EUDI Reference Implementation vs.\ Procivis One Core), each Verifier application returns results in different formats with different request parameters, even within the same ecosystem. While OID4VP standardizes the southbound interface between wallets and Verifiers, it does not address the northbound Verifier-to-organization interface, leaving each implementation free to define its own API. This contrasts with OIDC where a single integration works across all providers.

\subsection{interID}

%% CONSOLIDATED VERSION - Reduced from detailed architecture to brief summary + citation
%% Detailed descriptions removed as they're available in the published IEEE paper

interID, introduced in our previous work~\cite{yildiz2025}, addresses cross-ecosystem credential verification by providing a unified orchestration layer over multiple SSI framework-specific Verifier services. The system enables verification of credentials from Hyperledger Indy/Aries, EBSI, and EUDI ecosystems through a single integration point.

The three-layered architecture consists of: (1) a service layer integrating with framework-specific Verifiers (ACA-Py, Walt.id, EUDI Verifier), (2) a controller layer orchestrating verification and mapping unified proof templates to framework-specific formats, and (3) a presentation layer providing interfaces for administrators and Identity Holders. This design abstracts protocol complexities while maintaining ecosystem-specific features; Figure~\ref{fig:extended-architecture} in Section~\ref{design} depicts these interID capabilities in light blue alongside the extensions introduced in this work.

interID introduces \textit{proof templates} as a key abstraction by defining verification requirements once that apply across multiple ecosystems. When Verifiers initiate verification, interID presents wallet selection options, generates ecosystem-appropriate presentation requests, coordinates credential exchange, validates VPs, and returns results in JWT format. Our evaluation demonstrated successful verification across all supported ecosystems with minimal performance overhead.

\subsubsection{The Remaining Integration Gap}

While interID successfully addresses ecosystem-agnostic verification, it does not solve the organizational integration challenge. As an on-premises solution designed for self-hosted deployment, interID requires organizations to deploy and maintain their own verification infrastructure, including container orchestration, service configuration, and ongoing operational maintenance. Beyond these infrastructure complexities, organizations must still implement custom logic to integrate with interID's API (due to its non-standard interface), handle verification results, manage session correlation, and map verified claims to internal systems. The JWT-based result transmission lacks the standardization and security properties of established authentication protocols like OIDC.

This gap motivates our current work: providing an OIDC bridge that allows organizations to integrate SSI-based authentication using the same OIDC protocol they already support for federated identity, while delivering it as a SaaS offering that eliminates infrastructure deployment and maintenance requirements. Critically, this standardization provides flexibility. Organizations integrate once with OIDC and can subsequently switch between different Verifier applications without additional integration effort, just as they currently switch between different OIDC providers without fundamentally modifying OIDC client code.

\section{Related Work} \label{related}

The integration of SSI with conventional Identity and Access Management (IAM) protocols has emerged as a critical research area for enabling SSI adoption in existing enterprise infrastructure. We organize our review across three dimensions: SSI-OIDC bridge implementations, enterprise integration requirements, and multi-ecosystem interoperability, positioning our contributions against this landscape.

\subsection{SSI-OIDC Bridge Implementations}

Several research efforts bridge SSI with OIDC. \textbf{VC-AUTHN-OIDC}~\cite{curran2023vcauthnoidc}, developed by BC Gov and maintained by OpenWallet Foundation, pioneered OIDC bridges but remains tightly coupled to Hyperledger Aries, supports only AnonCreds/JSON-LD formats, lacks PKCE support (confidential clients only), and was not designed for multi-tenant SaaS deployment. This tight coupling reflects a deliberate design choice: by integrating directly with ACA-Py, VC-AUTHN-OIDC achieves production reliability within the Aries ecosystem at the cost of portability. Our architecture trades this tight integration for ecosystem independence through an adapter layer, accepting additional architectural complexity to support multiple verification backends.

\textbf{Hoops and Matthes}~\cite{hoops2024universal} advanced ecosystem-independence through nested OIDC (SIOPv2 + Core) with simple DIDs, but their prototype lacks security audit and PKCE support details remain unclear. Their nested OIDC approach offers an architectural advantage: by using SIOPv2 as the Verifier-facing protocol, it avoids implementing custom Verifier-specific adapters entirely. However, this design constrains claim richness to what SIOPv2 can convey and requires wallet support for the SIOPv2 protocol, limiting applicability to ecosystems where wallets already implement it. Earlier proof-of-concept work by \textbf{Lux et al.}~\cite{lux2020distributed} (Sovrin/Indy) and \textbf{Grüner et al.}~\cite{gruner2019integration} (uPort/Jolocom) demonstrated multi-platform principles but remain at prototype maturity with dated platform choices.

A concurrent 2024 work~\cite{ieee10646413} emphasizes established standards and custom domain-specific languages for policy configuration, but available details do not specify PKCE support, multi-tenancy design, or comprehensive ecosystem coverage.

\textbf{SAML Integration Alternative:} Yildiz et al.~\cite{yildiz2021connecting} demonstrated SSI integration via SAML for German higher education, showing that IdP-centric integration (where IdPs become Verifiers) addresses gradual SSI adoption within federations, complementing our RP-centric OIDC approach.

\subsection{Enterprise Integration Requirements and IAM Integration Surveys}

Systematic surveys establish theoretical foundations for SSI-IAM bridge design. Kuperberg and Klemens~\cite{kuperberg2022integration} surveyed over 40 SSI implementations across authentication protocols (OIDC, OAuth, LDAP, SAML, X.509), finding that only seven offered conventional IAM protocol integration, with merely two as open source. Their analysis identified three integration patterns: (A) IdPs modified to support SSI, (B) SSI Verifiers modified to support non-SSI protocols, and (C) OPs extended to support SSI-based authentication—our work exemplifies Pattern C.

Schardong and Custódio~\cite{schardong2022self} performed systematic literature review across 80+ papers, developing SSI taxonomy that identifies protocol integration with established IAM standards as critical for adoption. Their finding that protocol integration failure may jeopardize widespread adoption provides theoretical validation for our architectural approach.

A 2023 systematic review~\cite{bise2023iam} examined IAM requirements through design science methodology with expert validation, confirming that: (1) SSI integrates well with existing IAM infrastructure without wholesale replacement, (2) SSI solutions maintain strong similarity with OIDC-based approaches, and (3) claims from VPs integrate seamlessly into Access Tokens. This enterprise validation directly supports our bridge architecture strategy.

\subsection{Multi-Ecosystem Interoperability Challenges}

Grech et al.~\cite{grech2021blockchain} analyzed blockchain-based SSI in education, presenting four-dimensional interoperability framework: technical standards (W3C-VC, W3C-DID, W3C-JSON-LD), legal frameworks (eIDAS, GDPR), semantic models (Europass), and governance considerations. Their analysis of cross-border scenarios and EBSI infrastructure demonstrates why multi-ecosystem support is essential rather than optional, validating our architectural decision to support Aries, EBSI, and EUDI simultaneously.

\subsection{Positioning Our Contributions}

Our work addresses critical research gaps through three key innovations absent in prior implementations:

\textbf{Multi-Ecosystem Support:} We provide the first implementation explicitly supporting three major contemporary SSI ecosystems: Hyperledger Indy/Aries, EBSI, and EUDI while addressing cross-border European interoperability requirements and vendor independence concerns identified in surveys. As the acceptance of EUDI Wallet becomes mandatory across the EU~\cite{eidas2}, this architecture future-proofs Verifier investments.

\textbf{PKCE Implementation:} Explicit PKCE support (RFC 6819 best practices) enables public clients (single-page applications, native mobile apps) without requiring backend intermediaries. Prior works either explicitly omit PKCE (VC-AUTHN-OIDC) or leave implementation details unspecified (Hoops \& Matthes). This capability aligns with OAuth 2.1 evolution and eliminates deployment complexity requiring intermediary identity management systems.

\textbf{Multi-Tenancy for SaaS:} While prior implementations address multi-tenancy partially or not at all, our architecture enables SaaS deployment with explicit tenant isolation across all layers. This supports commercial models where organizations adopt SSI without deploying infrastructure, potentially accelerating mainstream adoption.

\textbf{Novel Security Analysis:} Beyond engineering contributions, we identify seven attack threats that emerge specifically at the intersection of SSI credential verification and multi-tenant OIDC bridge architectures, extending beyond RFC 6819's OAuth 2.0 threat model. Prior SSI-OIDC bridge works either omit security analysis entirely or address only standard OAuth 2.0 threats without considering SSI-specific and multi-tenant attack vectors.

Table~\ref{tab:comparison} synthesizes these differentiators against prior work. Our contributions combine multi-ecosystem support, modern security practices (PKCE), explicit SaaS architecture, and a systematic security analysis of novel bridge-specific threats, advancing practical viability for enterprise SSI adoption while maintaining full backward compatibility with existing OIDC infrastructure.

\begin{table*}[t]
\centering
\caption{Comparison of SSI-OIDC Bridge Implementations}
\label{tab:comparison}
\begin{tabular}{lp{2.2cm}p{2cm}p{1.5cm}p{2cm}p{2cm}}
\hline
\textbf{Work} & \textbf{Ecosystem Support} & \textbf{PKCE Support} & \textbf{Multi-Tenancy} & \textbf{Architecture} & \textbf{Maturity} \\
\hline
VC-AUTHN-OIDC~\cite{curran2023vcauthnoidc} & Aries/Indy only & No (confidential clients only) & Limited (via ACA-Py) & Coupled OIDC + ACA-Py & Production deployed \\
\hline
Lux et al.~\cite{lux2020distributed} & Sovrin/Indy only & No (implicit flow) & Not addressed & IdP-centric & Proof-of-concept \\
\hline
Hoops \& Matthes~\cite{hoops2024universal} & Ecosystem-agnostic (lightweight DIDs) & Authorization code (PKCE unclear) & Single tenant & Nested OIDC (SIOPv2 + Core) & Prototype (unaudited) \\
\hline
Grüner et al.~\cite{gruner2019integration} & Multi-platform (uPort, Jolocom) & Not addressed & Not addressed & Gateway + Trust Engine & Proof-of-concept \\
\hline
\textbf{Our Work} & \textbf{Aries, EBSI, EUDI} & \textbf{Yes (explicit)} & \textbf{Yes (SaaS)} & \textbf{Integration + IAM Cross-Layer} & \textbf{Production-oriented} \\
\hline
\end{tabular}
\end{table*}

This positioning demonstrates that our work addresses critical gaps in the SSI-OIDC bridge landscape identified through systematic surveys, providing the first, multi-ecosystem, PKCE-enabled, multi-tenant OIDC bridge implementation designed for production enterprise deployment. By combining support for major contemporary SSI ecosystems with modern security practices and explicit SaaS architecture, our work advances the practical viability of SSI adoption in enterprise environments while maintaining full backward compatibility with existing OIDC infrastructure.

\section{Concept and Design} \label{design}

To address this research and industry gap while maintaining the existing capabilities of interID, we design the SaaS and OIDC extension as modular components that extend the baseline architecture without modification.

\paragraph{Terminology:} Identity protocols assign different names to the organization consuming identity data: OIDC uses \textit{Relying Party}, SAML uses \textit{Service Provider}, and SSI uses \textit{Verifier}. Because our system bridges these protocols---a single integrating organization simultaneously acts as an OIDC RP and relies on SSI verification---we adopt the unified term \textit{client} throughout Sections~\ref{design}, \ref{implementation}, and~\ref{evaluation} to avoid ambiguity.

\subsection{Design Goals and Requirements}
Our design is guided by five essential requirements that address both technical integration challenges and practical deployment considerations.

\textbf{Backward Compatibility:} The OIDC bridge must function as an optional component that extends rather than replaces existing interID capabilities. This design recognizes the fundamental difference in use cases: while OIDC serves authentication and authorization purposes, VCs address a broader spectrum of proof requirements including diploma verification and driver's license validation. These proof scenarios do not require authentication but rather evidence of possession or qualification.

For example, a university application workflow demonstrates this distinction: a prospective student may use their identity credential for initial identification and authentication, but subsequently present a bachelor's degree as a VP to satisfy admission requirements for a master's program. The OIDC bridge handles only the authentication aspect, while the underlying interID infrastructure continues to support the full range of credential verification use cases independent of the authentication flow. This architectural separation ensures that clients requiring only credential verification do not incur OIDC-related complexity, while clients needing standardized authentication can leverage the OIDC bridge without losing access to advanced proof capabilities.

\textbf{Multi-Tenancy:} To enable SaaS deployment, the architecture must support multiple independent tenants with proper isolation of client credentials, proof templates, and user sessions. Each tenant operates as an autonomous instance with complete control over their verification requirements and integration configuration.

The system must provide tenants with self-service capabilities to define their verification logic through proof template creation, specifying which credential types, attributes, and validation rules their applications require. Tenants must be able to obtain OAuth 2.0 client credentials that authenticate their backend services when initiating authorization requests or proof verification flows. These credentials enable programmatic integration while maintaining strict isolation, ensuring that one tenant cannot access another tenant's proof templates, user sessions, or authentication state. The multi-tenancy architecture must support concurrent operation of numerous tenants without performance degradation or security boundary violations, providing each tenant with an isolated verification environment equivalent to operating their own dedicated interID instance.

\textbf{OIDC Protocol Compliance:} The system must implement OIDC specifications with sufficient fidelity that clients can integrate using standard OIDC flows without modifications. This includes supporting the authorization code flow with PKCE, providing properly formatted ID Tokens with standard claims, and exposing required discovery endpoints (.well-known/openid-configuration). The implementation must handle standard OIDC flows including authorization requests, token exchange, and session management according to specification requirements.

Additionally, according to the OIDC specification~\cite{oidcidtoken}, the standard claims set can be extended arbitrarily, and custom scopes can be defined, provided that the scope identifiers remain unique within the federation. This extensibility mechanism enables clients to request domain-specific attributes, such as document identifiers, professional certifications, or regulatory compliance status, by defining custom scopes that map to specific claims from proof template requirements. The bridge must be able to translate these custom scopes into appropriate proof requests directed to the underlying interID backend and incorporate the resulting credential attributes as additional claims in the issued ID Token, therefore, maintaining full OIDC compliance while supporting arbitrary claims not specified in the core OIDC specification. 

\textbf{Security Equivalence:} The bridge must provide security properties equivalent to traditional OPs, including Cross-Site Request Forgery (CSRF) protection through state parameters, replay attack prevention through nonce binding, authorization code security through PKCE, and secure session management isolating user authentication state. The implementation must resist common attack vectors applicable to OAuth 2.0 and OIDC flows, ensuring that the introduction of SSI-based verification does not compromise security guarantees expected from established IdPs.

\textbf{Modularity:} The system architecture must maintain clear separation between SSI verification logic (existing interID) and OIDC protocol handling (new bridge), enabling independent evolution and deployment of each component. The bridge component should interact with the interID backend through well-defined APIs without requiring modifications to the core verification engine. This separation allows organizations to utilize the bridge selectively based on their integration requirements, supports independent versioning and updates of each component, and prevents OIDC-specific concerns from propagating into the SSI verification.

\subsection{System Architecture Overview}

Building upon the original interID architecture described in Section~\ref{background} and our published work~\cite{yildiz2025}, we introduce two additional architectural layers that enable standardized authentication integration and multi-tenant SaaS deployment while preserving the modular design principles of the existing system.

\subsubsection{Integration Layer}

To enable standardized authentication protocol support, we introduce an \textbf{Integration Layer} positioned above the existing interID architecture. This layer exposes protocol-compliant interfaces that clients can integrate using standard OIDC client libraries, translating between established authentication protocols and the SSI verification capabilities provided by the underlying interID system.

The Integration Layer implements several critical functions. First, it provides standardized authentication endpoints conforming to widely-adopted protocol specifications, enabling RPs to integrate with interID using familiar authentication flows without implementing SSI-specific logic. Second, it coordinates protocol flows with SSI verification processes, translating protocol-specific requests into interID proof template invocations and synthesizing protocol-compliant responses from verification results. Finally, it generates standardized ID Tokens containing verified VP attributes as claims, enabling RPs to establish authenticated sessions based on cryptographically verified identity information.

\subsubsection{Identity and Access Management Platform Cross-layer}

To support multi-tenant SaaS deployment, we introduce an \textbf{IAM Platform} that provides cross-cutting services to all architectural layers. Unlike traditional architectural layers that process sequential request flows, the IAM Platform operates as a vertical service layer that multiple components invoke for authentication, authorization, and tenant context management.

The IAM Platform fulfills several essential functions across the architecture:

\begin{itemize}
\item \textbf{Presentation Layer Support:} The IAM Platform provides client admin authentication, enabling client administrators to securely access configuration interfaces with permissions scoped to their specific tenant. Administrators authenticate through challenge-response mechanisms, receiving Access Tokens that authorize operations on tenant-specific resources while enforcing strict isolation from other tenants' data.

\item \textbf{Integration Layer Support:} The IAM Platform validates client credentials when clients initiate authentication requests, ensuring that only registered clients with valid credentials can invoke the authentication endpoints.

\item \textbf{Inter-service Communication:} The IAM Platform issues service Access Tokens that enable components to invoke each other's APIs securely, with the Controller Layer validating these tokens before processing requests from the Integration Layer.
\end{itemize}

This centralized approach to authentication and authorization ensures consistent security policy enforcement across all architectural components and enables tenant isolation mechanisms.

\subsubsection{Controller Layer Extension}
We extend the Controller Layer to support client management, which handles the registration and configuration of clients. This enables clients to authenticate via OIDC or submit proof requests through interID's core capabilities. Client administrators manage these registrations via the admin interface.

\subsubsection{Extended Architecture Overview}

\begin{figure}[!htbp]
  \centering
  \includegraphics[width=\columnwidth]{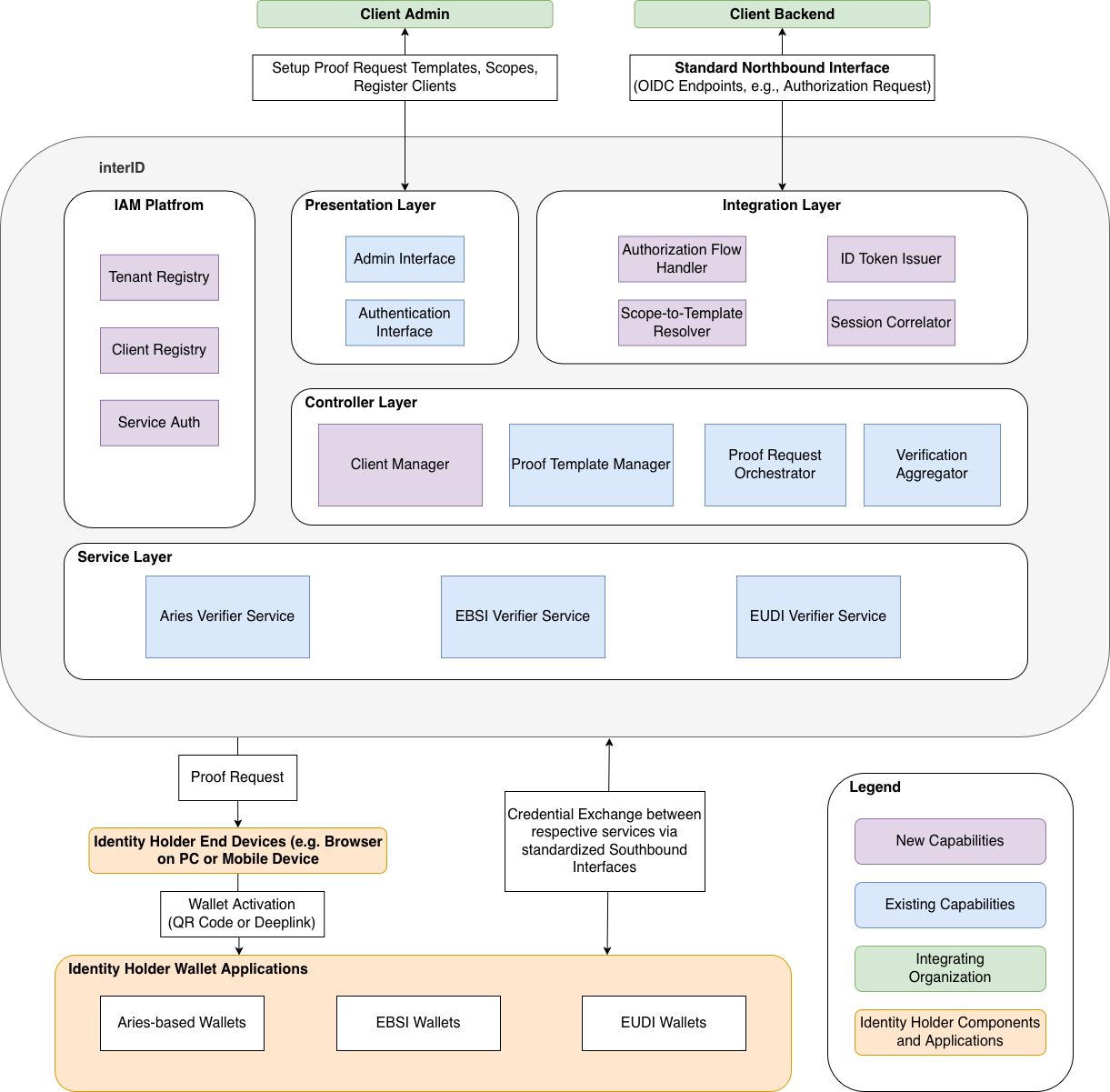}
  \caption{interID Extended Architecture}
  \Description{Architecture diagram illustrating the new Integration Layer and IAM Platform cross-layer, showing how the OIDC bridge extends the original interID architecture while maintaining modular design and preserving existing southbound interface interactions with SSI ecosystems}
  \label{fig:extended-architecture}
\end{figure}

Figure~\ref{fig:extended-architecture} illustrates the complete extended architecture. Clients interact with the Integration Layer through standardized protocol interfaces, remaining agnostic to the underlying SSI verification mechanisms. The Integration Layer coordinates with the IAM Platform for client authentication and with the Controller Layer for verification orchestration. The IAM Platform provides authentication and authorization services to the Presentation Layer for admin access, to the Integration Layer for client validation, and to inter-layer API calls for service authorization. The existing Controller and Service Layers continue to handle SSI verification without modification, maintaining ecosystem-specific verification logic independently. Identity Holders interact with the system through the Presentation Layer for wallet selection and credential presentation, with their wallets communicating directly with ecosystem-specific Verifier services to complete verification protocols.

\subsection{Functional Capabilities}

\begin{figure}[!htbp]
  \centering
  \includegraphics[width=\columnwidth]{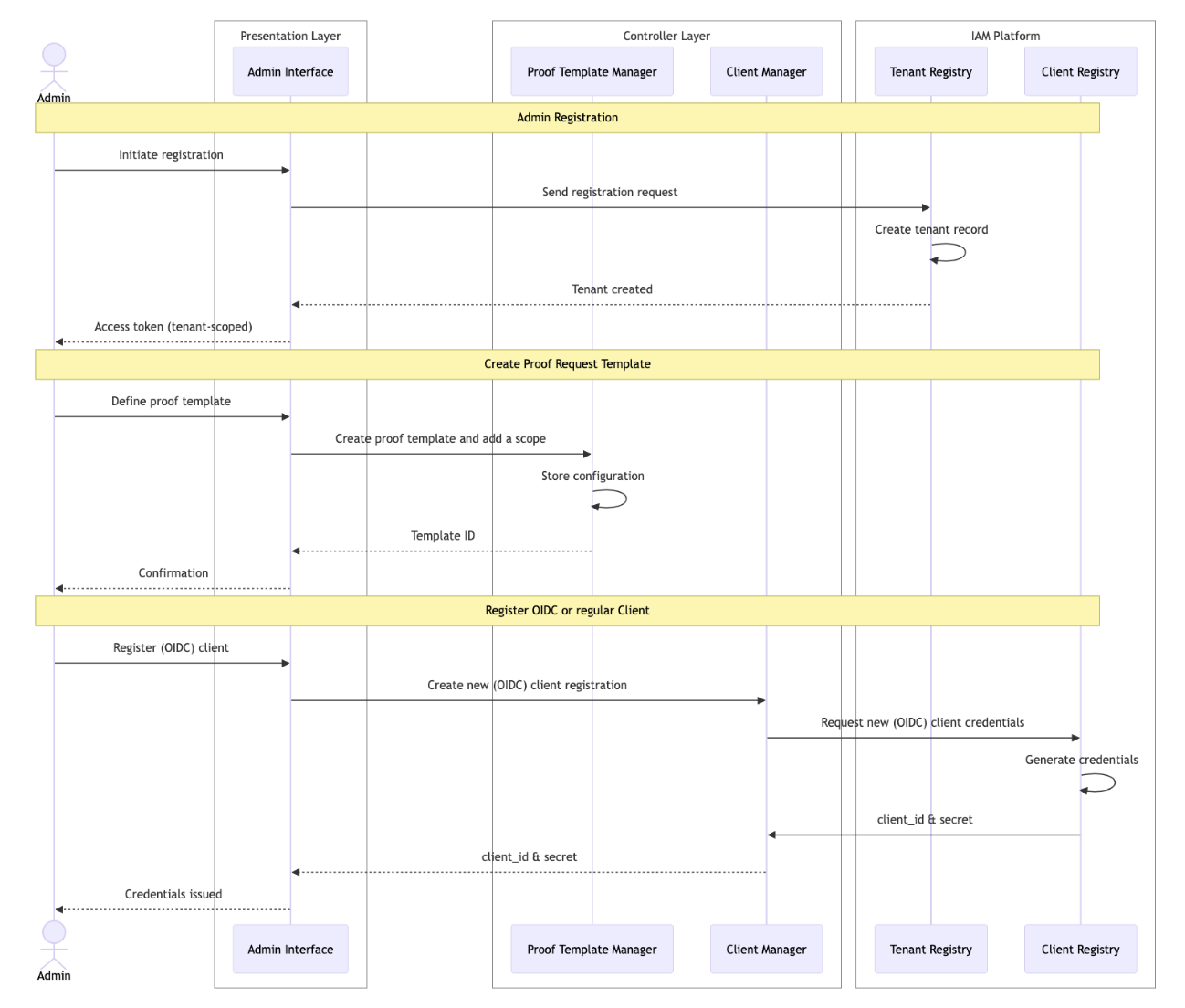}
  \caption{Client Admin Interactions: Tenant registration, proof template creation with scope and (OIDC) client registration}
  \Description{Sequence diagram showing three administrative workflows: tenant registration where administrators create organizational accounts, proof template creation where administrators define verification requirements and associate them with OIDC scopes, and OIDC client registration where administrators obtain OAuth 2.0 credentials for client applications}
  \label{fig:admin_functionalities}
\end{figure}

To realize the extended architecture and meet our design objectives, the following functional capabilities must be implemented across the logical architecture layers. Figure~\ref{fig:admin_functionalities} illustrates the administrative workflows (tenant registration, proof template creation, and client registration), while Figure~\ref{fig:auth_flow_concept} depicts the complete OIDC Authorization Code Flow with PKCE integrated with SSI credential verification.

\subsubsection{Admin Registration and Login}

Organizations adopting interID as a SaaS platform must first establish their tenant identity and administrative access. The admin registration capability enables organizational administrators to create tenant accounts and obtain authenticated access to tenant-specific configuration interfaces.

When an administrator initiates registration, the \textit{Tenant Registry} function creates a new tenant record with a unique tenant identifier. The administrator provides organizational information and credentials, which the IAM Platform validates and stores with appropriate isolation from other tenants. The IAM Platform enforces tenant-scoped authentication, ensuring that administrative credentials grant access only to the registering organization's resources and configuration interfaces.

Following successful registration, administrators access the \textit{Admin Interface} through the Presentation Layer. Upon login, the IAM Platform validates the administrator's credentials and issues an Access Token scoped to their tenant. This token authorizes all subsequent administrative operations within interID, including proof template creation, client registration, and configuration management.

\subsubsection{Proof Request Template Creation with OIDC Capabilities}

To leverage interID's verification capabilities, administrators define proof templates that specify which credentials and attributes their applications require. The extended proof request creation capability enables administrators to design verification requirements and map them to OIDC scopes, enabling authentication requests to automatically resolve to appropriate verification logic.

Administrators access the Admin Interface to create proof templates, specifying required credential types, attribute constraints, Issuer requirements, and validation rules. For multi-ecosystem deployments, administrators create ecosystem-specific proof request templates for each framework the organization wishes to support (such as EUDI, EBSI, Aries/Indy). Each proof request template transforms the unified proof template specification into framework-specific presentation request formats that the Service Layer can submit to ecosystem-specific Verifier services.

The key innovation enabling OIDC integration is the association of custom scopes with proof templates. When creating or updating proof templates, administrators can define custom scope identifiers that map to the unified proof template and its ecosystem-specific variants. For example, an administrator might create a proof template requesting government-issued identity credentials and associate it with a scope identifier such as \texttt{government\_identity}. When a client initiates an OIDC authentication request with this scope, the \textit{Scope-to-Template Resolver} within the Integration Layer translates the scope identifier to the corresponding proof template identifier and ecosystem-specific proof request templates.

Administrators also configure claim mapping rules as part of the proof request template definition, specifying how verified credential claims should map to OIDC claims in the ID Token. These mappings enable the ID Token Issuer to transform verified credentials into standardized OIDC claim structures that clients can consume using standard libraries.

\subsubsection{OIDC Client Registration}

Clients integrating with interID via OIDC require OAuth 2.0 credentials and configuration that enable them to initiate standardized authentication flows. The \textit{Client Manager} allows administrators to register clients and obtain credentials for OIDC integration.

Client administrators access the Admin Interface to register OIDC clients on behalf of the client. During registration, administrators specify the RP's callback URLs where the authorization endpoint will redirect after credential verification completes. The \textit{Client Registry} component stores this OIDC client registration along with the tenant identifier, ensuring that only this tenant's administrators can manage the registration.

Upon successful registration, the \textit{Client Registry} issues OAuth 2.0 client credentials consisting of a client identifier and client secret. These credentials uniquely identify the RP and enable programmatic authentication when exchanging authorization codes for ID Tokens. The IAM Platform stores credentials with tenant-scoped isolation, ensuring that the client can only access resources and proof templates associated with the registering tenant.

\subsubsection{Client Registration}

Beyond OIDC integration, organizations may require direct access to interID's proof verification capabilities for non-OIDC use cases, such as standalone proof requests or custom application flows. The client registration capability enables administrators to register applications and services that access interID through programmatic APIs.

Similar to OIDC client registration, administrators access the Admin Interface to register non-OIDC clients. During registration, administrators specify the client's purpose and which proof templates the client should be authorized to invoke. The Client Registry stores this registration along with the tenant identifier and issued credentials that the client presents when accessing interID's APIs.

The IAM Platform associates service credentials with specific tenants and authorized proof templates. When a registered client invokes interID's API endpoints, the Service Auth component validates the presented credentials against the Client Registry, ensures the client is authorized for the requested proof template, and issues a service Access Token that enables downstream components to process the request.

This capability maintains consistency with OIDC client registration while accommodating direct programmatic access patterns that do not require standardized authentication protocol compliance.

\subsubsection{Authentication Request with PKCE}

The authentication request with PKCE capability represents the primary integration point between clients and interID's OIDC bridge. This capability enables clients to initiate credential verification flows using standard OIDC authorization requests, leveraging interID's SSI verification capabilities while maintaining full OIDC protocol compliance.

When a user attempts to access a protected resource, the client's frontend constructs an OIDC authorization request including the client identifier, requested scopes, callback URL, state parameter for CSRF protection, and PKCE challenge parameters for authorization code security. The client's frontend redirects the user to the Integration Layer's authorization endpoint.

The \textit{Auth Flow Handler} component within the Integration Layer validates the authorization request, verifying client registration, callback URL matching, scope definitions, and PKCE challenge requirements. Upon successful validation, the Auth Flow Handler creates an OIDC session storing protocol parameters and generates a session identifier for correlation. The user is then redirected to the Authentication Interface for wallet selection and credential presentation.

The authentication process proceeds through the existing interID verification mechanisms. Upon successful verification, the Auth Flow Handler generates an authorization code bound to the PKCE challenge and redirects the user to the RP's callback URL with the authorization code and state parameter.

The client's backend exchanges the authorization code for tokens by contacting the token endpoint, presenting the authorization code, PKCE verifier, client identifier, and client secret. The Auth Flow Handler validates these parameters, and upon successful validation, the \textit{ID Token Issuer} generates a JWT-encoded ID Token containing claims derived from verified credentials according to the proof template's claim mapping rules. This ID Token is returned to the client's backend along with an Access Token.

%This capability demonstrates how interID transparently integrates SSI credential verification into standard OIDC flows, enabling clients to establish authenticated sessions based on cryptographically verified identity information using familiar OAuth 2.0 integration patterns. Section~\ref{implementation} describes the technical implementation of these components in detail.

\begin{figure*}[!htbp]
  \centering
  \includegraphics[width=\textwidth]{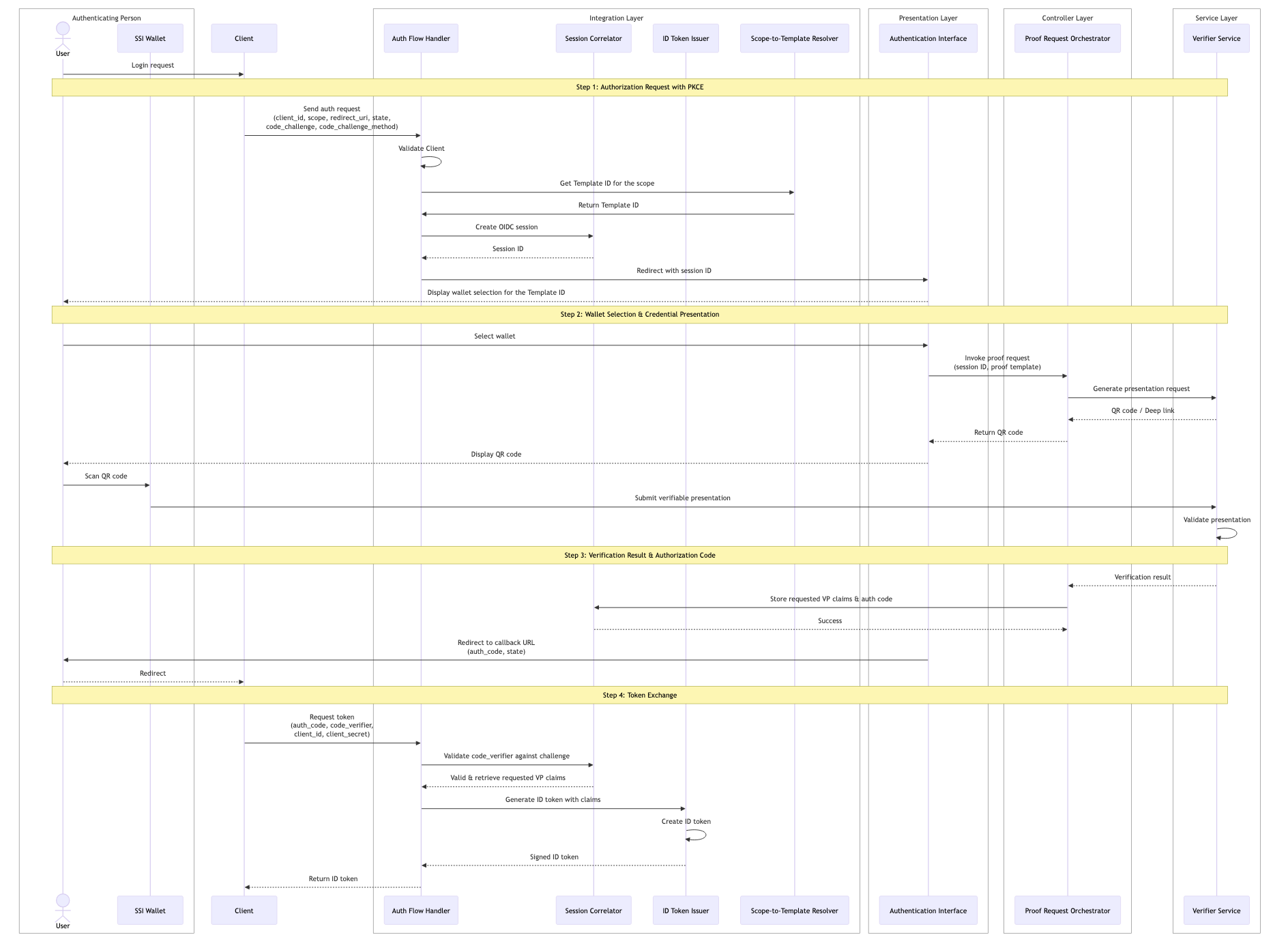}
  \caption{Authentication via Authorization Code flow with PKCE}
  \Description{Sequence diagram showing the complete OIDC Authorization Code Flow with PKCE integrated with SSI credential verification, including authorization request, wallet selection, credential presentation, verification by ecosystem-specific Verifier services, authorization code issuance, token exchange with PKCE verification, and ID Token generation with verified credential claims}
  \label{fig:auth_flow_concept}
\end{figure*}

As depicted in Figure~\ref{fig:auth_flow_concept}, the complete flow combines OIDC-conformant authorization code exchange with VC-based challenge-response and scope mapping.

\section{Implementation}\label{implementation}

This section describes the technical realization of the architectural design presented in Section~\ref{design}. The system extends the existing interID platform through containerized microservices preserving modularity and backward compatibility. Table~\ref{tab:component-mapping} maps architectural components to implementation services and data stores.

\begin{table*}[htbp]
\centering
\caption{Architecture-to-Implementation Component Mapping}
\label{tab:component-mapping}
\small
\begin{tabular}{|p{0.18\linewidth}|p{0.22\linewidth}|p{0.22\linewidth}|p{0.28\linewidth}|}
\hline
\textbf{Layer} & \textbf{Architectural Component} & \textbf{Implementation Service} & \textbf{Technology / Notes} \\
\hline
IAM Platform & Tenant Registry & Keycloak & MongoDB collection with tenant metadata \\
\cline{2-4}
& Client Registry & Keycloak + interID-backend & Keycloak: OIDC credentials; MongoDB: proof template associations \\
\cline{2-4}
& Service Auth & Keycloak & Service accounts for inter-service communication \\
\hline
Presentation & Admin Interface & admin-frontend & React/TypeScript, tenant configuration UI \\
\cline{2-4}
& Authentication Interface & auth-frontend & React/TypeScript, wallet selection and QR generation \\
\hline
Integration & Authorization Flow Handler & interID-oidc-backend & Express.js endpoints: /authorize, /token \\
\cline{2-4}
& ID Token Issuer & interID-oidc-backend & JWT signing with RS256 \\
\cline{2-4}
& Scope-to-Template Resolver & interID-oidc-backend & Queries MongoDB via interID-backend API \\
\cline{2-4}
& Session Correlator & interID-oidc-backend & Redis-backed session management \\
\hline
Controller & Client Manager & interID-backend & Client registration and credential management \\
\cline{2-4}
& Proof Template Manager & interID-backend & ProofTemplateController, MongoDB persistence \\
\cline{2-4}
& Proof Request Orchestrator & interID-backend & ProofRequestController, ecosystem selection \\
\cline{2-4}
& Verification Aggregator & interID-backend & Result normalization from Verifier services \\
\hline
Service & Aries Verifier Service & ACA-Py + interID-backend adapter & Docker container, AnonCredQueryBuilder \\
\cline{2-4}
& EBSI Verifier Service & Walt.id + interID-backend adapter & Docker container, WaltQueryBuilder \\
\cline{2-4}
& EUDI Verifier Service & EUDI Verifier + interID-backend adapter & Docker container, PIDQueryBuilder \\
\hline
Data & Session Store & Redis & OIDC sessions, auth tokens, authorization codes \\
\cline{2-4}
& Configuration Store & MongoDB & Proof templates, client configs, tenant data \\
\hline
\end{tabular}
\end{table*}

The implementation consists of two backend services, two frontend applications, three infrastructure services (Keycloak, Redis, MongoDB), and three ecosystem-specific Verifier containers, built with Node.js/TypeScript (Express.js) and React. \textbf{interID-backend} implements the Controller and Service Layers from the original interID architecture~\cite{yildiz2025}, exposing REST APIs for proof template management, proof request orchestration, and verification result aggregation through adapter patterns that transform unified templates into Verifier-specific formats. \textbf{interID-oidc-backend} implements the Integration Layer, orchestrating OIDC authorization requests by resolving scopes to proof templates, managing sessions in Redis, and generating signed JWT ID Tokens. \textbf{Keycloak} serves as the IAM Platform with a dedicated realm where each tenant registers and subsequently creates client accounts. Redis stores ephemeral session state with namespace isolation per tenant, while MongoDB stores persistent configuration with tenant-scoped access controls.

\subsection{Service Implementations}

\subsubsection{Main Backend Service (interID-backend)}

The main backend service remains largely unchanged from~\cite{yildiz2025}, preserving backward compatibility. The key extension for OIDC support is the addition of \texttt{scopes} arrays and \texttt{isAuthOnly} flags to proof templates, enabling scope-to-template mapping. The ProofTemplateController manages templates in MongoDB with ecosystem-specific format configurations (AnonCreds, JSON-LD, DIF Presentation Exchange, EBSI, EUDI), while specialized query builders (AnonCredQueryBuilder, WaltQueryBuilder, PIDQueryBuilder) transform unified templates into Verifier-specific formats.

A new Client Management Extension implements OIDC client lifecycle management following OAuth 2.0~\cite{rfc6749} and OpenID Connect~\cite{oidccore} requirements, synchronizing client data between MongoDB (proof template associations) and Keycloak (credential validation).

\subsubsection{OIDC Authentication Server (interID-oidc-backend)}

The OIDC backend implements a lightweight OP with four endpoints. The \textbf{Authorization Endpoint} (\texttt{/authorize}) validates client credentials against Keycloak, resolves requested scopes to proof templates, and redirects users to auth-frontend for credential presentation. The \textbf{Token Endpoint} (\texttt{/token}) performs PKCE verification, retrieves verification results from Redis, maps verified credential attributes to OIDC claims, and returns RS256-signed ID Tokens. We chose RS256 over ES256 for broad compatibility with existing OIDC client libraries, as RS256 remains the most widely supported algorithm across enterprise IAM tooling. The \textbf{Discovery} and \textbf{JWKS Endpoints} provide standard OIDC metadata and public keys for token validation.

\subsubsection{Authentication Frontend (auth-frontend)}

The authentication frontend provides Identity Holders with interfaces for credential presentation, implementing both same-device and cross-device flows.

Upon authentication initiation, the frontend retrieves the template to determine supported SSI ecosystems and dynamically renders wallet choices. For \textbf{same-device flows}, the system generates ecosystem-specific deep links (e.g., \texttt{eudi-openid4vp://} for EUDI, \texttt{didcomm://} for Aries). For \textbf{cross-device flows}, the system generates QR codes encoding presentation requests, employing URL shortening when necessary for QR code capacity limitations.

Following credential presentation, the frontend polls the backend for verification status using correlation identifiers. Upon successful verification, the frontend retrieves the authorization code and redirects users to the client's redirect URI, completing the OIDC authorization flow.

\subsection{Security Implementation}

The implementation incorporates security mechanisms at multiple architectural layers through cryptographic operations, session management, access control enforcement, and multi-tenant isolation. This section presents the technical implementation details; Section~\ref{sec:security-analysis} provides comprehensive security analysis and threat mitigation assessment.

\subsubsection{OAuth 2.0 and OIDC Protocol Security}

\textbf{PKCE Implementation:} The authorization endpoint mandates PKCE for all requests, validating \texttt{code\_challenge} parameters with SHA-256 hashing and 256-bit entropy code verifiers. Token endpoint validation:

\begin{lstlisting}[language=JavaScript,caption={PKCE Verification at Token Endpoint}]
const session = await redis.get(`AUTH_CODE:${code}`);
const computed = crypto.createHash('sha256')
  .update(codeVerifier).digest('base64url');
if (computed !== session.code_challenge) {
  throw new Error('PKCE verification failed');
}
\end{lstlisting}

Authorization codes expire after 10 minutes with single-use semantics enforced through Redis key deletion.

Beyond PKCE, the system enforces cryptographic binding at multiple levels. State parameters use 128-bit entropy UUIDs generated via \texttt{crypto.randomUUID()} to prevent CSRF attacks, while nonces embedded in ID Tokens prevent token replay. Challenge nonces for SSI verification sessions ensure freshness of credential presentations. ID Tokens are signed using RS256 with 2048-bit RSA keys that rotate every 90 days, with token audiences (\texttt{aud} claim) specifying intended client IDs under strict validation.

Session management implements defense-in-depth through httpOnly flags preventing JavaScript access, Secure flags ensuring HTTPS-only transmission, and SameSite policies (Lax for development, None for production). Session identifiers use UUIDv4 (122 bits entropy) with Redis expiration after 30 minutes. All credentials are time-limited: authorization codes expire after 10 minutes, authentication tokens after 5 minutes, and ID Tokens after 60 minutes, with Redis enforcing TTL-based expiration to bound the attack window for any compromised credential.

\subsubsection{Multi-Tenant Isolation}

\textbf{Keycloak-Based Client Authentication:} Each tenant registers distinct OIDC clients within Keycloak. The express-oauth2-jwt-bearer middleware validates bearer tokens and populates \texttt{req.auth.payload.sub} with the tenant ID, enabling tenant-scoped resource access throughout the system.

\textbf{Controller-Level Access Control:} Tenant isolation is enforced through explicit \texttt{userId} filters in all resource queries:

\begin{lstlisting}[language=JavaScript,caption={Tenant-Scoped Template Retrieval}]
list = async (req: Request, res: Response) => {
  const { auth } = req;
  if (!auth?.payload.sub) {
    return res.status(StatusCodes.UNAUTHORIZED)
      .json({ message: 'Unauthorized User' });
  }
  
  const result = await ProofTemplateModel.getMany({
    sortBy: sortBy as string,
    order: order as 'asc' | 'desc',
    limit: limit as number,
    page: page as number,
    condition: { userId: auth.payload.sub }
  });
  // ...
};
\end{lstlisting}

The MongoDB schema reinforces tenant isolation at the data model level through a mandatory \texttt{userId} field on all proof templates. Each template record includes a unique identifier, the owning tenant's user identifier, associated OIDC scopes, an authentication-only flag, and optional ecosystem-specific configurations for AnonCreds, JSON-LD, DIF Presentation Exchange, EBSI, and EUDI formats. This schema design ensures that all database queries are inherently tenant-scoped, as the \texttt{userId} condition in Listing~2 propagates through every data access path.

Client validation endpoints reinforce tenant ownership by cross-referencing tenant identifiers before authorizing operations, rejecting mismatches with \texttt{invalid\_client} errors.

Session data isolation leverages Redis key prefixing with distinct namespaces (\texttt{SESSION:}, \texttt{AuthCode:}) and TTLs per session type. We chose Redis key prefixing over separate Redis instances per tenant as it provides equivalent namespace isolation with lower operational complexity, avoiding per-tenant infrastructure provisioning while maintaining strict data boundaries. The implementation maintains three session types: OIDC Sessions (30-minute TTL) for authorization request parameters, Authentication Tokens (5-minute TTL) for template metadata and correlation identifiers, and Authorization Codes (10-minute TTL) as single-use tokens deleted upon first use. ID Tokens are generated as RS256-signed JWTs with standard OIDC claims and tenant-specific audience binding, with verified credential attributes included as additional claims.

\subsection{Verifier Service Integration}

InterID integrates with three containerized Verifier services---ACA-Py (Hyperledger Aries), Walt.id (EBSI), and EUDI Verifier---through adapter patterns implementing a common interface. Each adapter transforms unified proof templates into service-specific request formats and normalizes verification responses into consistent payloads, enabling uniform processing regardless of the underlying Verifier service or credential format. Adding support for a new ecosystem requires implementing a new adapter conforming to this interface without modifying the OIDC bridge or existing adapters.

\section{Evaluation} \label{evaluation}

We evaluate our system across three critical dimensions: security resilience against known attack vectors, inherent limitations of using SSI credentials for authentication purposes, and integration convenience compared to direct SSI implementation. The evaluation demonstrates that the bridge maintains security properties equivalent to established IdPs while substantially reducing integration complexity, though fundamental constraints in SSI credential design impose requirements on authentication use cases.

\subsection{Security and Threat Analysis}
\label{sec:security-analysis}

We analyze the security posture of interID through comprehensive threat modeling, identifying threats across three security domains: OAuth 2.0/OIDC protocol security, SSI credential verification security, and multi-tenant isolation in shared infrastructure. We identify 11 attack threats (AT.1-AT.11) and 9 implementation security controls (IS.1-IS.9). Table~\ref{tab:threats} summarizes our threat taxonomy and countermeasures, demonstrating defense-in-depth through overlapping control assignments. Our analysis demonstrates that the architecture provides security design equivalent to production IdPs while addressing novel SSI-specific and multi-tenant threats not covered by traditional OAuth 2.0 threat models.

\begin{table}[htbp]
\caption{Threat-Control Traceability Matrix}
\label{tab:threats}
\small
\begin{tabular}{|l|p{2.5cm}|p{2cm}|p{2cm}|}
\hline
\textbf{ID} & \textbf{Threat} & \textbf{Primary Control} & \textbf{Supporting Controls} \\
\hline
AT.1 & Auth Code Intercept & IS.1 (PKCE) & IS.2, IS.5 \\
AT.2 & CSRF & IS.2 (State) & IS.5, IS.8 \\
AT.3 & Token Replay & IS.2 (Nonce) & IS.3, IS.5 \\
AT.4 & Session Hijacking & IS.5 (Cookies) & IS.2, IS.9 \\
AT.5 & Request Manip. & IS.4 (Templates) & IS.3, IS.8 \\
AT.6 & Verification Spoof & IS.3 (Crypto) & IS.6 \\
AT.7 & Credential Replay & IS.2 (Challenge) & IS.3, IS.9 \\
AT.8 & Credential Isol. & IS.6 (Keycloak) & IS.4, IS.7 \\
AT.9 & Session Isolation & IS.7 (Redis NS) & IS.6 \\
AT.10 & Template Isol. & IS.4 (Filters) & IS.6, IS.7 \\
AT.11 & Token Misuse & IS.3 (Audience) & IS.6 \\
\hline
\end{tabular}
\end{table}

\subsubsection{Threat Model}

We identify 11 attack threats mapped to the trust boundaries in our extended architecture (Figure~\ref{fig:extended-architecture}). OAuth 2.0/OIDC threats (AT.1-AT.4) target the protocol communication between clients and the Integration Layer, as documented in RFC 6819~\cite{rfc6819}. SSI-specific threats (AT.5-AT.7) emerge in the verification process spanning the Integration Layer, Controller Layer, and Service Layer interactions with Identity Holder wallets. Multi-tenant isolation threats (AT.8-AT.11) address tenant boundary violations that can occur across all architectural layers.

\textbf{Authorization Code Interception (AT.1):} Attackers may intercept authorization codes during redirect flows. The mandatory PKCE implementation (Section~\ref{implementation}, OAuth 2.0 and OIDC Protocol Security) with SHA-256 challenge provides cryptographic binding. Code verifiers use 256 bits of entropy, making brute-force computationally infeasible. Authorization codes expire after 10 minutes with single-use semantics. \emph{[Mitigated by: IS.1, IS.2, IS.5]}

\textbf{Cross-Site Request Forgery (AT.2):} CSRF attacks attempt to trick users into unauthorized authorization requests. OAuth state parameters bind to user sessions through cryptographically random UUIDs (Section~\ref{implementation}, OAuth 2.0 and OIDC Protocol Security), validated on return. Redirect URI validation uses exact string matching against Keycloak pre-registered values. \emph{[Mitigated by: IS.2, IS.5, IS.8]}

\textbf{Token Replay and Theft (AT.3):} ID Token replay attacks are prevented through nonce binding (Section~\ref{implementation}, OAuth 2.0 and OIDC Protocol Security). The bridge generates cryptographically random nonces for each request, embeds them in ID Tokens, and validates they match the initiating session. Token audience (\texttt{aud}) claims specify intended client IDs, preventing token reuse across clients. \emph{[Mitigated by: IS.2, IS.3, IS.5]}

\textbf{Session Hijacking (AT.4):} Session management (Section~\ref{implementation}, OAuth 2.0 and OIDC Protocol Security) incorporates httpOnly flags preventing JavaScript access, Secure flags ensuring HTTPS transmission, and SameSite=Lax policies preventing CSRF. Session identifiers use UUIDv4 (122 bits entropy). Redis sessions expire after 30 minutes with automatic cleanup. \emph{[Mitigated by: IS.5, IS.2, IS.9]}

\textbf{Proof Request Manipulation (AT.5):} The bridge enforces proof template integrity by storing templates in MongoDB with tenant-scoped access control (Section~\ref{implementation}, Multi-Tenant Isolation) and generating proof requests server-side. Templates are retrieved using \texttt{userId} filters, and template identifiers in authentication tokens are cryptographically signed using RS256. \emph{[Mitigated by: IS.4, IS.3, IS.8]}

\textbf{Verification Result Spoofing (AT.6):} The architecture implements end-to-end cryptographic validation. VPs undergo cryptographic verification by ecosystem-specific Verifier services (Section~\ref{implementation}, Verifier Service Integration) before results reach the bridge. The bridge accepts verification results only from authenticated Verifier services through internal APIs with service-to-service OAuth tokens. \emph{[Mitigated by: IS.3, IS.6]}

\textbf{Credential Presentation Replay (AT.7):} The interID backend generates unique challenge nonces (UUIDv4, 128-bit entropy) for each verification session, which wallets must include in presentations. Verifier services validate expected challenge values, and each presentation is processed exactly once. Time-limited authentication tokens (5-minute expiration) constrain the replay attack window. \emph{[Mitigated by: IS.2, IS.3, IS.9]}

\textbf{Client Credential Isolation (AT.8):} Keycloak-based client management (Section~\ref{implementation}, Multi-Tenant Isolation) provides cryptographic isolation. Each client tenant registers as a distinct OIDC client with unique credentials validated before processing requests. Client-specific scope configurations limit proof templates accessible to each tenant. \emph{[Mitigated by: IS.6, IS.4, IS.7]}

\textbf{Session Data Isolation (AT.9):} Redis session keys incorporate client identifiers (\texttt{SESSION:\{clientId\}:\{sessionId\}}), providing namespace isolation. Session retrieval logic validates that the requesting client matches the session owner. See Section~\ref{implementation} for implementation details. \emph{[Mitigated by: IS.7, IS.6]}

\textbf{Proof Template Isolation (AT.10):} MongoDB queries (Section~\ref{implementation}, Multi-Tenant Isolation) include mandatory \texttt{userId} filters implemented through tenant-aware middleware. Database queries are parameterized to prevent NoSQL injection. \emph{[Mitigated by: IS.4, IS.6, IS.7]}

\textbf{Token Audience Validation (AT.11):} Generated ID Tokens include \texttt{aud} claims specifying intended client IDs. Validation logic strictly verifies audience matching, preventing token reuse across clients within the same or different tenants. See Section~\ref{implementation} for token generation implementation. \emph{[Mitigated by: IS.3, IS.6]}

\subsubsection{Implementation Security Controls}

The implementation incorporates nine security controls organized across multiple architectural layers, implementing defense-in-depth where each attack threat is addressed by two to three independent controls (Table~\ref{tab:threats}).

At the protocol level, PKCE (IS.1) cryptographically binds authorization requests to token exchanges with 256-bit entropy code verifiers, directly mitigating authorization code interception. Cryptographic binding parameters (IS.2) provide broader protection through state parameters (128-bit UUIDs) that prevent CSRF, nonces that prevent token replay, and challenge nonces that prevent credential presentation replay, collectively addressing five of the eleven identified threats. Cryptographic validation (IS.3) ensures end-to-end integrity through RS256-signed ID Tokens, ecosystem-specific credential verification by Verifier services, and strict audience validation.

At the application level, server-side template enforcement (IS.4) stores templates with tenant-scoped access through mandatory \texttt{userId} filters and parameterized queries, preventing both template manipulation and cross-tenant access. Secure session management (IS.5) combines httpOnly cookies, Secure flags, SameSite policies, UUIDv4 identifiers, and 30-minute expiration to protect session integrity. Input validation (IS.8) applies Zod schema validation, exact redirect URI matching, and scope validation before template resolution.

At the infrastructure level, multi-tenant authorization (IS.6) leverages Keycloak-based tenant isolation with service-to-service OAuth tokens and minimal-privilege service accounts, addressing the broadest range of multi-tenant threats. Data isolation (IS.7) enforces Redis namespace isolation and mandatory tenant filters at the data access layer. Deployment security (IS.9) provides containerized process isolation, network policies limiting lateral movement, and aggressive token expiration (authorization codes 10 minutes, authentication tokens 5 minutes, ID Tokens 60 minutes). In production deployments, these application-level isolation mechanisms would be complemented by hypervisor-level and infrastructure-level isolation provided by the hosting platform.

\subsubsection{Security Evaluation}

Our implementation addresses authorization code flow threats from RFC 6819~\cite{rfc6819} (AT.1--AT.4) through controls IS.1--IS.5. We extend this baseline with seven additional threats (AT.5--AT.11) that fall outside RFC 6819's scope and represent novel contributions of our security analysis. Threats AT.5--AT.7 target the SSI verification layer---proof request manipulation, verification result spoofing, and credential presentation replay---which are attack vectors unique to credential-based authentication and not addressed by traditional OAuth 2.0 threat models. These are mitigated through controls IS.2--IS.4 and IS.6. Threats AT.8--AT.11 address multi-tenant isolation boundaries, including credential isolation, session data isolation, template isolation, and token audience validation, which are SaaS-specific vectors not present in single-deployment models. These are mitigated through controls IS.4, IS.6, and IS.7 enforced across all architectural layers.

It is important to distinguish between threats inherent to SSI systems and those specific to our bridge architecture. Credential presentation replay (AT.7) and verification result integrity (AT.6) are challenges common to all SSI Verifier deployments; our contribution lies in demonstrating that introducing the OIDC bridge layer does not weaken the underlying security guarantees. The multi-tenant threats (AT.8--AT.11), by contrast, are novel to our SaaS architecture and require purpose-built isolation mechanisms.

Compared to alternative SSI-OIDC bridges, our implementation addresses three critical security gaps: PKCE support enables public clients without backend intermediaries, comprehensive multi-tenant isolation prevents cross-tenant attacks in shared infrastructure, and modular separation between the OIDC bridge and SSI verification engine reduces the attack surface compared to monolithic architectures. We note that this analysis is design-level; implementation-level validation through independent penetration testing is discussed in Section~\ref{future}.

\subsection{Limitations and Constraints}

While the OIDC bridge significantly reduces integration complexity, fundamental characteristics of SSI credential design impose constraints on authentication use cases that differ from traditional IdPs.

\subsubsection{Unique Identifier Requirements}

SSI-based authentication requires credentials to contain stable, unique identifiers that persist across presentations. However, privacy-preserving credential design often intentionally avoids persistent identifiers to prevent correlation tracking.

Authentication systems typically rely on stable identifiers for user account lookups and session correlation. SSI credentials may provide identifiers through: (1) explicit identity claims (document numbers, email addresses), (2) cryptographic key binding (public keys as pseudonymous identifiers), (3) (SD-)JWTs with consistent \texttt{sub} claims~\cite{sdjwt}, or (4) ephemeral identifiers (changing with each presentation).

\textbf{AnonCreds Correlation Challenges:} Hyperledger Indy AnonCreds present unique challenges due to their privacy-preserving design. AnonCreds use zero-knowledge proofs that enable holders to prove credential possession without revealing linkable identifiers~\cite{anoncreds}. For AnonCreds-based authentication to function, proof requests must explicitly request a stable identifying claim (e.g., \texttt{email}, \texttt{documentNumber}) that the holder discloses as a revealed attribute. This transforms authentication into attribute verification where the authenticating identifier is a revealed claim rather than an implicit credential property.

The EUDI Architecture and Reference Framework addresses this limitation through a two-phase approach~\cite{eudi_arf_1.6}. After initial identification using PID credentials, organizations can issue dedicated Strong User Authentication (SUA) attestations to users. These SUA attestations contain persistent pseudonymous identifiers bound to the user's identity, enabling subsequent authentication without requiring full PID presentation. This approach maintains regulatory compliance while providing the stable identifiers necessary for session management and account correlation.

\subsubsection{Absence of UserInfo Endpoint}

Standard OIDC implementations provide a UserInfo endpoint that allows RPs to retrieve user claims after authentication by presenting an Access Token. This endpoint enables separation between authentication (ID Token) and attribute retrieval (UserInfo request), supporting use cases where RPs may need to fetch updated user attributes during an active session.

The interID OIDC bridge deliberately omits the UserInfo endpoint to comply with eIDAS 2.0 regulatory requirements~\cite{eudi_arf_1.6}. The regulation prohibits intermediaries acting on behalf of Verifiers from storing identity data obtained from credential presentations. Since implementing a UserInfo endpoint would require the bridge to retain user attributes between the initial authentication and subsequent UserInfo requests, this capability conflicts with the regulatory framework governing EUDI Wallet interactions.

This design choice enforces a privacy-preserving architecture where all requested user claims must be included in the ID Token during initial authentication. The bridge operates as a stateless intermediary that facilitates credential presentation and verification without persisting identity attributes. Organizations integrating with the bridge must configure proof templates to request all necessary claims upfront, as subsequent attribute retrieval through a UserInfo endpoint is not supported.

While this constraint differs from traditional OIDC providers, it aligns with the broader SSI principle of minimizing intermediary data retention and provides stronger privacy guarantees. Clients requiring dynamic attribute updates during active sessions must implement alternative mechanisms, such as requesting fresh authentication with updated proof templates or maintaining separate processes for attribute refresh that initiate new verification flows.

\subsubsection{Session Persistence and Account Recovery}

Traditional IdPs offer password reset mechanisms and account recovery flows that SSI-based authentication cannot easily replicate due to decentralized credential storage.

\textbf{Device Loss Scenarios:} When users lose devices containing digital wallets, credential access is lost with no centralized recovery mechanism. Some wallet implementations provide backup mechanisms using seed phrases or cloud backup, but these are wallet-specific rather than protocol-level features. Organizations deploying SSI-based authentication must provide alternative access mechanisms for users experiencing credential loss.

\textbf{Credential Lifecycle Management:} Credentials have finite validity periods and may require renewal. When credentials expire, users lose authentication capability until obtaining renewed credentials from Issuers. Traditional IdPs handle updates transparently, while SSI users must actively obtain renewed credentials.

\textbf{Multi-Device Access:} Traditional IdPs enable authentication from any device by entering credentials. SSI-based authentication typically requires credentials to be present on the authentication device, complicating multi-device scenarios. In cross-device flows, users must use their device containing the wallet while authenticating on another device (e.g., scanning a QR code on a desktop computer with their smartphone wallet) which is more complex than simply typing username and password on the same device. Users must either manage such cross-device flows or maintain separate credentials for each device (and wallet).

\subsubsection{Trust Framework Dependencies}

SSI authentication relies on cryptographic verification of credential authenticity, which depends on trust frameworks defining which Issuers are trusted for which credential types.

\textbf{Issuer Trust Configuration:} Clients must configure trust frameworks specifying which Issuers are acceptable. This requires governance decisions about Issuer eligibility, credential schema requirements, and trust level mappings. The complexity of this configuration varies significantly across SSI ecosystems. The EUDI framework and EBSI provide comprehensive solutions through trusted lists and the EBSI ledger as root of trust, minimizing client configuration burden~\cite{eudi_arf_1.6,ebsi-trusted-lists}. Other frameworks like Hyperledger Indy require more extensive manual configuration, where client administrators must explicitly specify trusted Issuers while creating Indy-specific proof request templates.

\textbf{Schema and Claim Variations:} Different Issuers may issue credentials with different schemas for conceptually equivalent identity attributes. For example, multiple government agencies might issue identity credentials with variations in claim names, date formats, or semantic meanings. The EUDI framework and EBSI address this challenge through standardized attestation rule books and mandatory schema compliance enforced through trust infrastructure~\cite{eudi_arf_1.6}. Clients requiring interoperability across multiple Issuers in less standardized ecosystems may need to implement claim normalization logic or maintain multiple proof templates.

\subsubsection{Regulatory and Compliance Considerations}

Organizations subject to regulatory requirements must carefully evaluate whether SSI-based authentication meets compliance obligations.

\textbf{Audit and Logging:} Regulations often require detailed audit trails of authentication events. The interID bridge provides comprehensive logging of verification events, but the decentralized nature of SSI means the bridge cannot log credential usage across different Verifiers.

\textbf{Authentication Assurance Levels:} Regulatory frameworks like NIST SP 800-63 define authentication assurance levels (AAL) based on authentication strength~\cite{nist80063}. Mapping SSI-based authentication to these frameworks is ecosystem-specific. The EUDI framework provides clear mechanisms including wallet unit attestation, device binding, and authentication protocol specifications that map to defined assurance levels~\cite{eudi_arf_1.6}. Other frameworks present more challenges where Levels of Assurance (LoAs) are not clearly defined, and wallet implementations may or may not fulfill specific regulatory requirements without explicit certification programs.

\subsection{Integration Complexity Analysis}

Having established the security posture and inherent limitations of SSI-based authentication, we now evaluate the practical integration burden that our bridge architecture alleviates. Organizations seeking to integrate SSI-based credential verification face a fundamental challenge: the lack of standardized northbound interfaces between Verifier applications and backend systems. While Verifier applications successfully abstract southbound complexity (handling ecosystem-specific protocols and cryptographic verification), they expose disparate, vendor-specific APIs to consuming organizations. 

Even within the same ecosystem, different Verifier implementations (e.g., EUDI Reference Implementation vs. Procivis One Core) expose different APIs, response formats, and integration patterns. This forces organizations to implement custom wrappers that construct requests to vendor APIs, handle asynchronous verification through vendor-specific callback mechanisms or polling patterns, parse vendor-specific response formats where verification results may be indicated through different JSON structures (\texttt{verified: true} vs. \texttt{valid: true} vs. nested proof objects), and map heterogeneous data structures to internal identity models.

The interID SaaS offering with OIDC bridge addresses this integration complexity by providing a standardized northbound interface that abstracts both Verifier vendor diversity and credential format heterogeneity. We evaluate this integration advantage through qualitative task comparison, as no standardized baseline for direct multi-ecosystem SSI integration exists against which to measure development time. Instead, we demonstrate the structural complexity reduction through two representative use cases using task comparison tables, and validate the OIDC-side integration effort through a reference Relying Party implementation (Section~\ref{sec:ref-rp}).

\subsubsection{Use Case 1: Identity Proofing (KYC)}

Financial institutions commonly rely on Identity Verification (IDnV) providers such as Signicat, IDnow, and Onfido for KYC compliance. Some IDnV providers have adopted OIDC-based integration patterns~\cite{signicat-oidc}, enabling organizations to integrate electronic identity verification through standard OIDC flows. Organizations have built substantial infrastructure around these patterns: standard OIDC client libraries, JWT validation utilities, session management infrastructure, and compliance databases with audit logging.

The OIDC bridge enables substituting centralized IDnV services with decentralized credential verification while preserving most established integration patterns. Table~\ref{tab:kyc-comparison} compares the integration tasks required for each approach.

\begin{table}[htbp]
\centering
\caption{Integration Task Comparison: KYC Use Case}
\label{tab:kyc-comparison}
\small
\begin{tabular}{|p{0.45\linewidth}|p{0.45\linewidth}|}
\hline
\textbf{Direct SSI Integration} & \textbf{OIDC Bridge Integration} \\
\hline
\multicolumn{2}{|c|}{\textit{Initial Setup}} \\
\hline
Deploy Verifier services (ACA-Py, Walt.id, EUDI Verifier, Procivis containers) & 
Register OIDC client credentials \\
\hline
Configure vendor-specific service parameters (varies even within same ecosystem) & 
Configure custom scopes for proof templates \\
\hline
Set up container orchestration, networking, monitoring & 
Update JWKS endpoint configuration \\
\hline
\multicolumn{2}{|c|}{\textit{Development Tasks}} \\
\hline
Implement vendor-specific wrapper (different endpoints per vendor) & 
\textit{Reuse existing OIDC authorization flow} \\
\hline
Build proof request UI for wallet selection & 
\textit{Handled by auth-frontend} \\
\hline
Implement session correlation or adapt to vendor callback mechanisms & 
\textit{Standard OIDC state/nonce handling} \\
\hline
Parse vendor-specific response formats (\texttt{verified: true} vs. \texttt{valid: true} vs. proof objects) & 
Parse custom claims from standard JWT structure \\
\hline
Extract credential attributes from vendor-specific or ecosystem-specific structures & 
Extract attributes from consistent JWT claim structure \\
\hline
Map credential attributes to internal schemas & 
Map JWT claims to internal schemas \\
\hline
Build vendor-specific error handling & 
\textit{Standard OIDC error responses} \\
\hline
\multicolumn{2}{|c|}{\textit{Operational Tasks (On-Prem vs SaaS)}} \\
\hline
Monitor Verifier service health & 
\textit{No infrastructure monitoring} \\
\hline
Manage vendor-specific dependencies and updates & 
\textit{No dependency management} \\
\hline
Handle Verifier service scaling and availability & 
\textit{SaaS handles scaling} \\
\hline
\multicolumn{2}{|c|}{\textit{Multi-Vendor Complexity}} \\
\hline
Repeat integration for each vendor (limited code reuse even within ecosystem) & 
Single OIDC integration works across all vendors/ecosystems \\
\hline
\end{tabular}
\end{table}

As shown in Table~\ref{tab:kyc-comparison}, direct SSI integration requires implementing vendor-specific wrappers, parsing vendor-specific response formats, and handling vendor-specific operational characteristics. Critically, organizations trust the Verifier application's validation results but must implement custom parsing logic for each vendor's response format, as vendors indicate verification success through different JSON structures and key names. 

OIDC bridge integration abstracts this vendor diversity through standard OIDC flows and consistent JWT ID Token structure, requiring only configuration changes and minor claim extraction for domain-specific attributes (e.g., \texttt{document\_number}, \texttt{issuing\_authority}) that appear consistently in JWT claims regardless of the underlying Verifier vendor or ecosystem.

\subsubsection{Use Case 2: Single Sign-On (SSO)}

Organizations deploy SSO systems using standardized OIDC protocols across application portfolios, representing substantial investments in client libraries, user provisioning workflows, and authorization policy engines. The OIDC bridge enables extending SSO with VC-based authentication while preserving most existing infrastructure. Table~\ref{tab:sso-comparison} illustrates the integration tasks for each approach.

\begin{table}[htbp]
\centering
\caption{Integration Task Comparison: SSO Use Case}
\label{tab:sso-comparison}
\small
\begin{tabular}{|p{0.45\linewidth}|p{0.45\linewidth}|}
\hline
\textbf{Direct SSI Integration (Per Application)} & 
\textbf{OIDC Bridge Integration} \\
\hline
\multicolumn{2}{|c|}{\textit{SSO Provider Configuration}} \\
\hline
Deploy Verifier infrastructure & 
Register interID as OIDC provider (once) \\
\hline
Create custom wrapper for Verifier vendor & 
Define custom scopes for credential types \\
\hline
\multicolumn{2}{|c|}{\textit{Application Integration}} \\
\hline
Implement vendor-specific wrapper in each app & 
\textit{Reuse existing OIDC client} \\
\hline
Build proof request UI in each app & 
Add scope parameters to existing auth request \\
\hline
Implement session correlation or callbacks in each app & 
\textit{Standard OIDC state handling} \\
\hline
Parse vendor-specific verification results in each app & 
Extend claim extraction for custom attributes from ID Token \\
\hline
Handle vendor-specific errors in each app & 
\textit{Standard OIDC error handling} \\
\hline
\multicolumn{2}{|c|}{\textit{Infrastructure Reuse}} \\
\hline
Cannot reuse existing SSO infrastructure & 
Reuses OIDC flows, JWT validation, session mgmt. \\
\hline
Each app requires independent implementation & 
Centralized provider config, per-app claim extraction \\
\hline
\multicolumn{2}{|c|}{\textit{Vendor Changes}} \\
\hline
Update each app when switching Verifier vendors & 
Update OP configuration \\
\hline
Changes propagate across all applications & 
Changes isolated to SSO provider layer \\
\hline
\end{tabular}
\end{table}

Table~\ref{tab:sso-comparison} demonstrates the multiplier effect in SSO contexts. With direct SSI integration, each application requires independent implementation of vendor-specific wrapper, proof request handling, and response parsing. When organizations switch Verifier vendors (e.g., from EUDI Reference Implementation to Procivis One Core), changes propagate across all applications due to different APIs and response formats. 

With the OIDC bridge, organizations configure interID as an OP once and leverage existing OIDC client implementations across all applications, requiring only minor claim extraction extensions per application. Vendor changes affect only the centralized OIDC provider configuration, not individual applications. For organizations with dozens or hundreds of applications, this architectural difference substantially reduces both initial integration effort and ongoing maintenance burden.

\subsubsection{Reference RP Implementation}\label{sec:ref-rp}

To validate the integration claims in Tables~\ref{tab:kyc-comparison} and~\ref{tab:sso-comparison}, we implemented a reference Relying Party application that integrates with the OIDC bridge. The backend integration logic---comprising OIDC session initialization, PKCE code verifier and challenge generation, authorization URL construction, authorization code exchange, and JWKS-based ID Token signature verification---requires approximately 180 lines of TypeScript. The implementation uses exclusively standard libraries (Express, axios, jsonwebtoken, Node.js crypto module) with no SSI-specific dependencies. Configuration consists of standard OIDC parameters: issuer URL, client credentials, redirect URI, and requested scopes. The resulting integration is structurally identical to integrating any conventional OIDC provider; a developer familiar with OAuth 2.0 and OIDC can implement it without SSI domain knowledge. Notably, the same codebase contains a direct interID API integration path for non-OIDC use cases, which requires Keycloak service account authentication, interID-specific API calls, and custom token validation---additional complexity that the OIDC bridge entirely abstracts.

\subsubsection{Developer Experience}

The OIDC bridge substantially reduces SSI integration complexity by eliminating specialized knowledge requirements for vendor-specific APIs and response format parsing. Developers integrate using familiar OIDC authorization flows, standard JWT token handling, and established session management patterns, as demonstrated by the reference implementation above. While custom scopes and claims require minor extensions beyond typical OIDC integration, either through manual JWT claim parsing or lightweight library wrappers, this represents substantially less complexity than direct SSI integration.

Direct SSI integration confronts developers with sparse and vendor-specific documentation, where API semantics vary even among Verifiers supporting the same ecosystem. Organizations must understand vendor-specific endpoint structures, response format variations (different JSON keys indicating success/failure across vendors), callback or polling mechanisms for asynchronous verification, and ecosystem-specific concepts including credential format differences (AnonCreds vs. JSON-LD vs. mDoc), trust registry resolution mechanisms, and revocation checking procedures. 

Each vendor introduces distinct API patterns with limited knowledge transfer. For example, expertise integrating the EUDI Reference Implementation provides minimal advantage when integrating Procivis One Core despite both supporting EUDI credentials, as their northbound interfaces differ substantially.

This knowledge requirement differential has organizational implications. Direct integration requires hiring specialized SSI developers familiar with specific Verifier vendors or contracting system integrators offering multi-month and multi-million implementation projects. The OIDC bridge enables integration with existing team capabilities, as handling custom scopes and claims represents a straightforward extension of standard OIDC integration patterns that most web developers understand. Organizations avoid the build-versus-buy decision inherent in custom wrapper development, as the bridge represents predominantly standard OIDC integration with minor domain-specific extensions rather than complete custom wrappers. Validating these developer experience observations through controlled studies with representative developers remains future work.

\subsubsection{Multi-Ecosystem and Multi-Vendor Support}

The complexity reduction becomes more pronounced when organizations require support for multiple SSI ecosystems or need flexibility to switch Verifier vendors. Table~\ref{tab:multi-ecosystem} compares the additional work required to support additional ecosystems or change vendors.

\begin{table}[htbp]
\centering
\caption{Multi-Ecosystem and Vendor Flexibility Comparison}
\label{tab:multi-ecosystem}
\small
\begin{tabular}{|p{0.45\linewidth}|p{0.45\linewidth}|}
\hline
\textbf{Direct SSI Integration} & \textbf{OIDC Bridge Integration} \\
\hline
\multicolumn{2}{|c|}{\textit{Adding EUDI Support}} \\
\hline
Deploy EUDI Verifier service & 
\textit{Already supported by bridge} \\
\hline
Implement EUDI vendor-specific wrapper & 
Configure EUDI proof template \\
\hline
Parse EUDI vendor response format & 
\textit{Same JWT claim structure as before} \\
\hline
\multicolumn{2}{|c|}{\textit{Adding Aries Support}} \\
\hline
Deploy ACA-Py service & 
\textit{Already supported by bridge} \\
\hline
Implement Aries-specific wrapper & 
Configure Aries proof template \\
\hline
Parse ACA-Py response format & 
\textit{Same JWT claim structure as before} \\
\hline
\multicolumn{2}{|c|}{\textit{Switching EUDI Verifier Vendor}} \\
\hline
Implement new vendor wrapper & 
Update proof template configuration \\
\hline
Update response parsing for new vendor format & 
\textit{No client-side changes} \\
\hline
Test and deploy changes across all integration points & 
Test template configuration only \\
\hline
\multicolumn{2}{|c|}{\textit{Code Reuse}} \\
\hline
Minimal across ecosystems, limited within ecosystem across vendors & 
Complete (same OIDC integration) \\
\hline
\end{tabular}
\end{table}

Table~\ref{tab:multi-ecosystem} demonstrates that multi-ecosystem support with direct integration requires near-complete re-implementation for each ecosystem, and even switching vendors within the same ecosystem requires substantial API client and parsing logic changes. With the OIDC bridge, organizations configure additional proof templates while reusing the same OIDC integration code. 

This flexibility is particularly valuable as the SSI ecosystem matures. Organizations can evaluate different Verifier vendors, switch between them based on cost or feature considerations, or support multiple ecosystems for different use cases, all without modifying their application integration code. For large organizations requiring EUDI (European operations), Aries (North American and Asian partnerships), and EBSI (EU cross-border services), this represents the difference between three independent integration efforts versus one OIDC integration with three proof template configurations.

\subsubsection{Extensibility to Additional Ecosystems}

The adapter-based architecture is designed for extensibility beyond the three currently supported ecosystems. Adding support for a new SSI ecosystem requires implementing a query builder that transforms unified proof templates into ecosystem-specific request formats and a response normalizer that maps verification results into the common claim structure. The OIDC bridge layer and client-facing integration remain entirely unchanged, as the scope-to-template mapping abstracts ecosystem-specific details. This extensibility applies to any SSI ecosystem based on Verifiable Credential and Verifiable Presentation exchange patterns (e.g., did:web-based ecosystems, PolygonID). Ecosystems with fundamentally different interaction models, such as mDL verification over Bluetooth Low Energy without a network-accessible Verifier service, would require deeper architectural adaptation at the service layer to accommodate non-HTTP communication patterns.

\section{Future Work} \label{future}

The core contribution---an ecosystem-agnostic OIDC bridge with multi-tenant isolation, PKCE support, and scope-to-template mapping---is complete and evaluated. The extensions below fall into two categories: \textit{production hardening} (penetration testing, OIDC4IDA compliance, additional OIDC flows) encompasses security validation and protocol coverage improvements that strengthen the existing architecture, while \textit{scope extensions} (SAML support, eIDAS 2.0 regulatory conformance) address complementary protocol families and regulatory frameworks beyond the current OIDC focus.

\subsection{Security Validation Through Penetration Testing}

While our threat modeling (Section~\ref{sec:security-analysis}) identifies attack vectors with corresponding countermeasures, independent penetration testing would validate production readiness. Such testing should encompass OAuth 2.0 and OIDC attack patterns including authorization code interception with and without PKCE, CSRF through state manipulation, token replay across client contexts, and session hijacking, thereby validating implemented countermeasures (IS.1--IS.5) under adversarial conditions.

Multi-tenant isolation testing should include cross-tenant access attempts through database query manipulation, session hijacking across tenant boundaries, token misuse between tenants, and privilege escalation to verify architectural isolation mechanisms (IS.4, IS.6, IS.7). SSI-specific vectors such as proof request manipulation, verification result spoofing, credential presentation replay with reused nonces, and man-in-the-middle attacks on Verifier communication channels should be tested to validate SSI-specific controls (IS.2, IS.3, IS.6).

Finally, infrastructure security testing should address container escape attempts, network policy bypass, service authentication weaknesses, and cryptographic vulnerabilities including weak randomness, improper key storage, and inadequate TLS configuration. Independent security validation would identify implementation-specific vulnerabilities beyond design review, validate controls under realistic attack scenarios, and provide external validation for enterprise adoption and regulatory compliance.

\subsection{OpenID Connect for Identity Assurance Compliance}

Full compliance with OpenID Connect for Identity Assurance (OIDC4IDA)~\cite{openid-4-ida-spec} would further enhance interoperability with IDnV providers that support the \texttt{verified\_claims} structure and assurance metadata defined in the specification. This standardization would enable organizations to integrate both traditional IDnV services and SSI-based credential verification through a unified interface, providing consistent claim structures, evidence metadata, and assurance levels across verification providers.

\subsection{Additional OIDC Flows}

The current implementation supports the Authorization Code Flow with PKCE, which provides robust security for both public and confidential clients. Supporting the standard Authorization Code Flow without PKCE requirement would facilitate adoption by legacy enterprise systems and confidential clients that have not yet implemented PKCE. While PKCE is strongly recommended as a security best practice~\cite{oauth2bcp}, many existing OIDC client implementations, particularly in established enterprise environments, rely on client secret authentication alone. Providing this option would reduce migration barriers for organizations with mature IAM infrastructure while maintaining security through traditional client authentication mechanisms. The bridge could enforce PKCE for public clients while allowing confidential clients to authenticate without PKCE, aligning with current industry practices during the PKCE transition period.

\subsection{SAML Protocol Support}

Many enterprise applications and academic federations rely on SAML 2.0 and cannot easily migrate to OIDC. Implementing a SAML 2.0 IdP interface alongside the OIDC OP would enable the bridge to serve both protocol families simultaneously. SAML assertions would contain verified credential attributes as SAML attributes, mirroring the OIDC ID Token approach. This would enable SSI integration for organizations with SAML-based infrastructure, particularly in higher education (Shibboleth federations) and large enterprise environments where SAML remains the dominant protocol.

\subsection{Regulatory Conformance for eIDAS 2.0 Intermediary Role}

Operating as a multi-tenant SaaS platform positions interID within the eIDAS 2.0 intermediary role definition~\cite{eidas2}, requiring regulatory compliance in future implementations. Our research agenda encompasses several specific regulatory requirements. First, strict data minimization and ephemeral processing must be enforced through architectural constraints ensuring that interID processes data solely for forwarding requests and managing protocol complexity, with no data storage (including personal data, verifiable credentials, or metadata like IP addresses) beyond the immediate transaction lifecycle. Second, the platform must satisfy core relying party obligations as defined in eIDAS 2.0, specifically regarding data privacy principles, robust security measures, and full transparency to the end-user.

Additionally, a dedicated module must handle the mandatory registration of the interID platform within a Member State's framework, including mechanisms to identify and authenticate the platform to the end-user prior to any data retrieval request. The platform also requires a resilient logging and audit trail system that captures sufficient, non-personal data to ensure that every interaction is signed, timestamped, and auditable by relevant supervisory bodies, maintaining accountability without compromising privacy. Furthermore, the implementation must respect the eIDAS 2.0 mandate that relying parties (and intermediaries) cannot refuse the use of pseudonyms if the user's identification is not legally required for the service. Finally, validation mechanisms must utilize the EU Trusted Lists (EUTL) and the free-of-charge systems provided by Member States to verify the authenticity and validity of EUDI Wallets and electronic attestations of attributes in real-time.

\section{Conclusion} \label{conclusion}

This work addresses a critical barrier to SSI adoption: the integration complexity that forces organizations to implement ecosystem-specific and Verifier application-specific wrappers despite standardization efforts at the credential and protocol levels. By extending the interID ecosystem-agnostic verification platform with an OpenID Connect bridge, we provide organizations with a standardized northbound interface that abstracts both Verifier API diversity and credential format heterogeneity. Organizations can integrate SSI-based authentication using familiar OIDC flows, standard client libraries, and established development patterns, substantially reducing integration complexity compared to direct SSI implementation.

The multi-tenant architecture leveraging Keycloak for IAM enables SaaS deployment where multiple organizations share infrastructure while maintaining strict data isolation, security boundaries, and independent configuration. This deployment model reduces operational burden for adopting organizations while providing the flexibility to support diverse verification requirements through scope-to-proof-template mappings. The implementation demonstrates that SSI-based authentication can achieve security properties equivalent to established IdPs through comprehensive threat mitigation, including PKCE implementation, defense-in-depth multi-tenant isolation, and end-to-end cryptographic verification of credential authenticity.

However, fundamental characteristics of SSI credential design impose constraints on authentication use cases that differ from traditional IdPs. Privacy-preserving credential formats may lack persistent identifiers required for session correlation and account management, credential loss scenarios require alternative access mechanisms due to decentralized storage, and user experience patterns differ from established username/password flows. These limitations are not technical deficiencies but rather inherent properties of SSI's privacy-by-design philosophy, requiring organizations to carefully evaluate use case alignment and consider ecosystem-specific recommendations or requirements for repeated authentication.

The significance of this work extends beyond technical integration convenience. By providing organizations with a familiar, standardized integration path, the OIDC bridge addresses the chicken-and-egg adoption challenge that has hindered SSI deployment: organizations hesitate to invest in SSI integration due to complexity and fragmentation, while SSI ecosystems struggle to demonstrate adoption without widespread organizational support. The bridge enables organizations to adopt SSI-based authentication with similar effort to adding Microsoft, Google, or Okta as federated IdPs, potentially accelerating SSI transition from theoretical paradigm to practical authentication infrastructure.

This work contributes to SSI-OIDC bridge research through production-oriented design addressing PKCE for enhanced security, comprehensive multi-tenant isolation architecture, and validated support for three major SSI ecosystems (Aries/Indy, EBSI, and EUDI). The implementation demonstrates that bridging SSI with established authentication protocols is not merely possible but practical, with security, convenience, and deployability properties suitable for production use. By eliminating the integration barrier between SSI verification and consuming organizations, we aim to enable mainstream SSI adoption and accelerate transition toward user-controlled, cryptographically verifiable digital identity.

\bibliographystyle{ACM-Reference-Format}
\bibliography{sample-base}

\end{document}